\documentclass[useAMS,usenatbib]{mn2e}
\usepackage[dvips]{graphicx}

\def\apj{{\rm ApJ}}

\def\etal{{\rm et~al.\ }}

\def\simlt{\lower.5ex\hbox{$\; \buildrel < \over \sim \;$}}
\def\simgt{\lower.5ex\hbox{$\; \buildrel > \over \sim \;$}}

\newcommand{\uv}[1]{\hat{\mathbf{#1}}}

\title[LSS and Halo Alignments]{The Influence of Large--Scale
  Structure on Halo Shapes and Alignments}

\author[G. Altay \etal]{
Gabriel Altay$^{1}$\thanks{E-mail: galtay@andrew.cmu.edu},
J\"org M. Colberg$^{1,2}$ and
Rupert A.C. Croft$^{1}$\\
$^{1}$Carnegie Mellon University, Department of Physics, 5000 Forbes Ave,
Pittsburgh, PA 15213, USA\\ 
$^{2}$University of Pittsburgh, Department of Physics and Astronomy, 3941 O'Hara Street, 
Pittsburgh PA 15260, USA\\}

\begin{document}
\pagerange{\pageref{firstpage}--\pageref{lastpage}} \pubyear{2005}

\maketitle

\label{firstpage}

\begin{abstract}
Alignments of galaxy clusters (the Binggeli effect), as well
as of galaxies themselves have long
 been studied both observationally and theoretically.
Here we test the influence of large-scales structures 
and tidal fields on the shapes and alignments of
cluster--size and galaxy--size dark matter halos. We use a high--resolution 
N--body simulation  of a $\Lambda$CDM universe, together with the 
results of Colberg et al. (2005), who identified filaments connecting pairs of clusters.
We find that cluster pairs connected by a filament are strongly aligned with
the cluster-cluster axis, whereas unconnected
ones are not. For smaller, galaxy--size halos, there also is an alignment signal, but its 
strength  is independent of whether the halo is part of an obvious large--scale structure. 
Additionally, we find no measureable dependence of galaxy
halo shape on membership of a filament.
We also quantify the influence of tidal fields
and find that these do correlate strongly with alignments of halos. The
alignments of most halos are thus caused by tidal fields, with cluster--size
halos being strongly aligned through the added mechanism of infall of matter from
filaments. 
\end{abstract}
 

\section{Introduction}

Galaxy redshift surveys such as the 2dFGRS (Colless et al. 2001) or the Sloan Digital 
Sky Survey (York et al. 2000) and N--body simulations of cosmic structure formation (for
example Springel et al. 2005 and references therein) demonstrate the
existence of a complicated network of
matter. At the most prominent positions in this network, massive clusters of galaxies
can be found, which are interconnected by filaments and, to a lesser degree, sheets.
In both observations and simulations clusters are aspherical systems, a fact which has
lead to investigations of the degree and origin of asphericity and of possible alignments 
between neighbouring objects. In this paper we use an N--body simulation of a 
$\Lambda$CDM universe to study the link between this network of structure and the 
shapes and alignments of galaxies and clusters.

Binggeli (1982) first investigated the alignment of galaxy clusters, finding that 
for the 44 Abell clusters in a sample there was a strong longitudinal alignment signal
(clusters tend to point towards each other, the ``Binggeli effect''). This was seen
for cluster-cluster separations
of up to $\sim$ 15 $h^{-1}$ Mpc\footnote{Throughout this work, we express the Hubble 
constant as $H_0 = 100\,h\,$km/sec/Mpc.}. Most follow--up studies, optical and otherwise,
have confirmed
Binggeli's results (Flin 1987, Rhee \& Katgert 1987, West 1989, Rhee et al. 1992, Plionis 1994, 
West et al. 1995, Chambers et al. 2000 and 2002), although there were also some negative 
reports (Struble \& Peebles 1985, Ulmer et al. 1989).

On the theoretical side, much attention has been devoted to the shapes of dark matter halos 
(for example, the radial density profile was examined by Navarro et al. 1997, Moore et al. 1999,
and the asphericity of galaxies and/or clusters by Jing \& Suto 2002, Bailin \& Steinmetz 2004, 
Hopkins et al. 2004, Kasun \& Evrard 2004, Allgood et al. 2005, Lee et al. 2005b, 
and Paz et al. 2005). The alignments of halos have been getting somewhat less attention. 
Splinter et al. (1997), 
Onuora \& Thomas (2000), and Faltenbacher et al. (2002) have all found significant alignments of  
galaxy cluster halos, as did Kasun \& Evrard (2004), Hopkins et al. (2005), and Basilakos et al. (2005). 
Galaxy--sized halos also have a strong tendency to be aligned in the same direction
as other nearby halos (see e.g. Heavens, Refregier \& Heymans 2000, Croft \& Metzler 2000),
as well as pointing along the direction vector to nearby halos (e.g., Li \& Croft 2005). This
latter signal, the intrinsic density--shear correlation
has been recently seen in observational galaxy data by Mandelbaum \etal (2005), and 
Agustsson \& Brainerd (2005).

It is not immediately obvious what causes the alignment of halos. As shown in Van Haarlem \&
Van de Weygaert (1993), clusters tend to orient themselves toward the direction of the last
matter infall (as shown in Colberg et al. 1999, matter falls into cluster predominantly from
filaments). But there is also a positive correlation between the inertia tensor of a cluster
and its surrounding tidal field (Bond et al. 1996; as shown in Lee et
al. 2005b, the axis--ratio distribution of halos can be modelled analytically on the basis of this, also
see Lee et al. 2005a). In reality, both effects will probably be intertwined. 

Hopkins et al. (2005) attempt to connect the alignment between cluster pairs and large--scale structure by
looking at the number density of clusters contained in a cylinder that connects the two clusters. They 
find that as the number density rises, so does the average cluster alignment. While this indicates
that filaments between clusters might cause increased alignment, for a more general analysis it
is necessary to investigate cluster samples for which the inter-cluster filaments are
found with reference to the density field itself. Such a set of filaments would also make
possible an investigation of shapes and alignments of halos much smaller
than galaxy clusters.

Colberg et al. (2005) investigated inter--cluster configurations of matter in
a  high--resolution simulation of cosmic structure  and found a
complete set of filaments. Here we will
use their data as the basis for a detailed investigation of the connection between large--scale
structure and halo alignments. In particular, we will investigate whether there is a
connection between the alignment of pairs of clusters and the existence (or non--existence) of
a filament between them. We will also study whether halos of mass smaller than
that of a massive cluster are aligned with filaments and/or the tidal fields of the clusters.
The latter is interesting in the light of an algorithm proposed by Pimbblet (2005) to locate filaments
in galaxy redshift catalogues.

This work is organized as follows. In Section~\ref{simulation} we describe the simulation, in 
Section~\ref{alignments} we study the alignments of halos between pairs of clusters connected
by filaments or with voids in between them, and in Section~\ref{ellip}, we briefly re--visit
ellipticities of halos. Section~\ref{summary} contains a summary and discussion.

\begin{figure}
 \includegraphics[angle=-90,width=85mm]{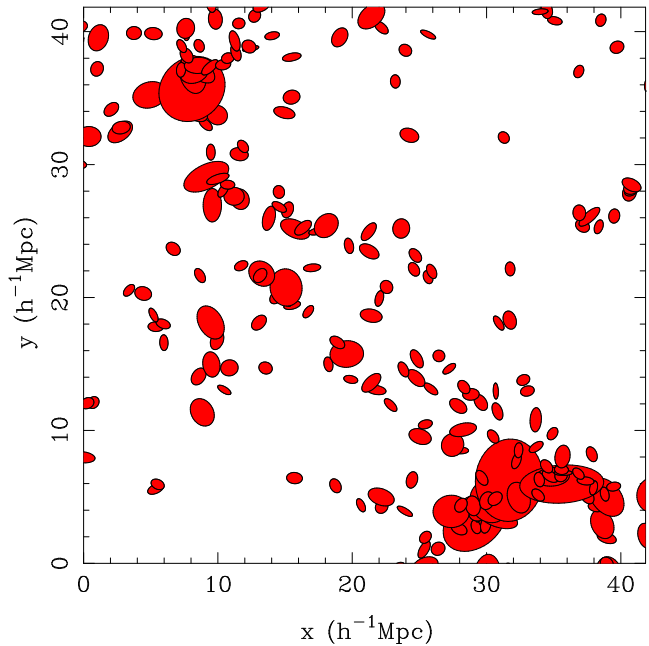}
  \caption{A part of a thin slice of thickness 15\,$h^{-1}$\,Mpc through the simulation volume,
           showing all halos as ellipsoids. The ellipsoids are scaled according to their mass,
           and their orientations correspond to those of the halo mass distributions.
           The plot is centered on one of the filaments found by Colberg \etal (2005), which
           spans two clusters, in the top left and bottom right hand corners of the panel.}
  \label{fig:image}
\end{figure}

\section{The Simulation} \label{simulation}

We make use of the high--resolution $\Lambda$CDM simulation introduced in Jenkins et al. (1998).
The simulation parameters ($\Omega$ = 0.3, $\Lambda$ = 0.7, $h$ = 0.7, and $\sigma_8$ = 0.9) 
are in good agreement with the currently accepted standard cosmology (Colless et al. 2001, Spergel 
et al. 2003, Seljak et al. 2005). The simulation follows the evolution of 256$^3$ Dark Matter 
particles in a cubical volume of size (141.3\,$h^{-1}$\,Mpc)$^3$ on a side, resulting in a 
particle mass of $1.4 \cdot 10^{10}\,h^{-1\,}M_{\sun}$. 

\subsection{The Halo Catalog}

The group catalogue is obtained by running a standard friends--of--friends group finder on the
full particle set, using a linking length of $b=0.2$ times the mean interparticle separation. 
We use all groups with 50 particles or more for the analysis, which results in a total sample of 
17461 halos. Our choice of the minimum halo mass is motivated by the fact that for less than
50 particles, the structure of a halo cannot be reliably determined
(see e.g. the tests in Kasun \& Evrard 2004). In the following, we will 
refer to the 170 most massive groups as clusters and to all others as halos. The cluster sample 
was designed to match the space density of Abell clusters as outlined in Colberg et al. (2005).

For each halo, we compute its principal axes by diagonalizing the moment of inertia tensor 
\begin{equation}
I_{ij} = \sum x_i x_j\,,
\end{equation}
where the sum is over all particles in the halo, and the coordinates are defined with respect to 
the center of mass of the group. The resulting eigenvalues $a$, $b$, and $c$ are sorted by size,
in descending order. The ellipticity of a group is then defined by
\begin{equation}
\epsilon = 1 - \sqrt{c/a}\,.
\end{equation}

\subsection{Filament Finding}

We make use of the filament catalogue described by Colberg et al. (2005). In that
paper, intercluster 
filaments were found by investigating the configuration of  matter between neighbouring clusters. The mass
distributions between all pairs of clusters (up to the 12th nearest neighbour of
each cluster) were projected onto orthogonal
planes. The results were then visually classified according to the
appearance of the projections (essentially into either a filament,
absence of a structure, or rarely, a sheet). We will use the Colberg \etal (2005) classifications to examine alignments of
halos and clusters with the cluster--cluster axes, and how it depends on the 
presence of a filaments..

Figure~\ref{fig:image} shows a small piece of the simulation volume,
with halos being plotted as ellipsoids. The ellipsoids are scaled according to their mass,
and the orientations of the ellipsoids correspond to those of the actual mass distributions. The region
shown in the plot encompasses a pair of clusters connected by a filament. The same
filament is shown bottom--center
in Figure~1 of Colberg et al. (2005). The filament is clearly visible, although any coherent alignments
of the halos is difficult to pick out by eye. We will examine this statistically in the next section.

\begin{figure*}
  \begin{center}
    \begin{tabular}{cc}
      \begin{minipage}{64mm}
        \begin{center}
          \includegraphics[width=64mm]{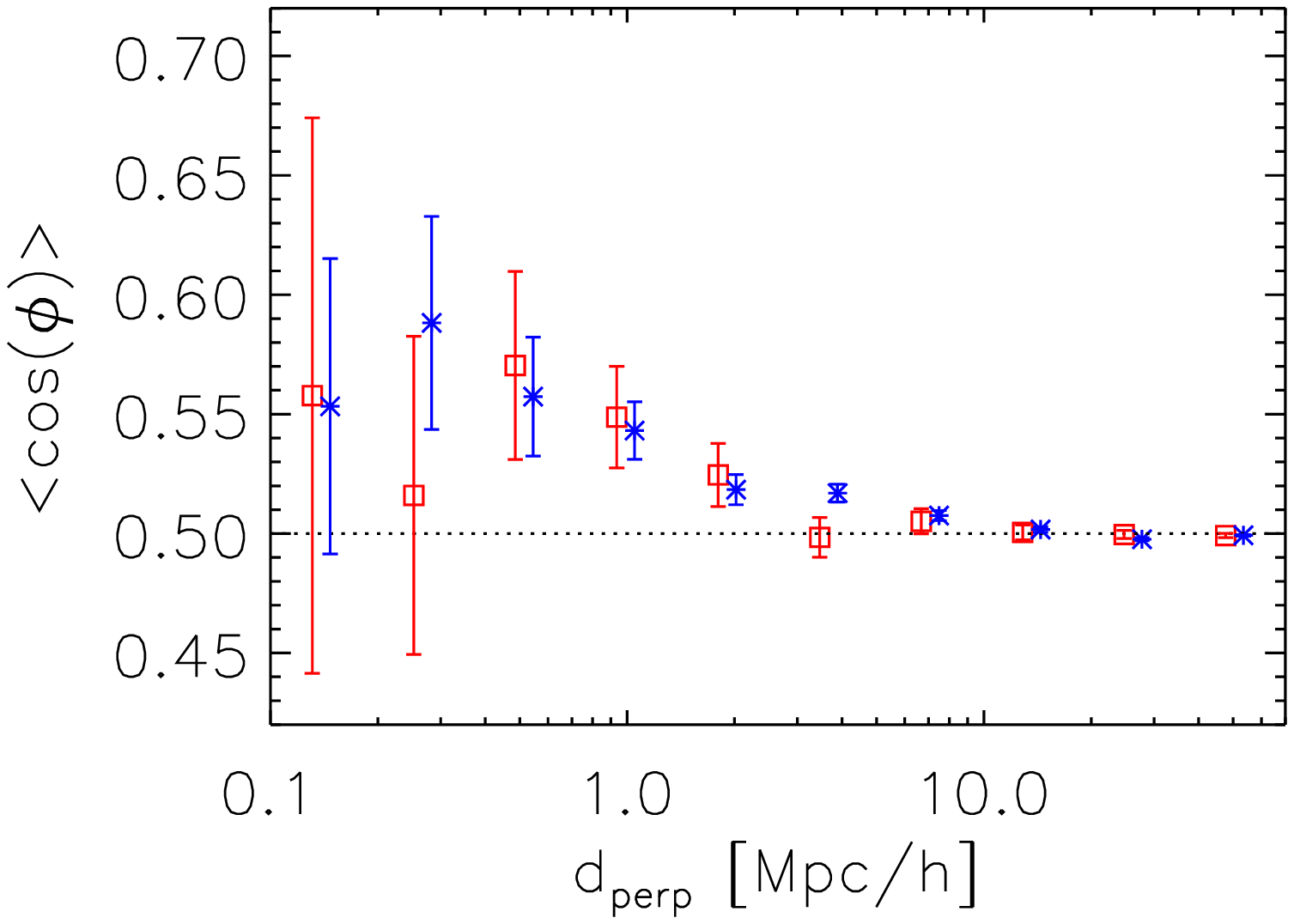}
        \end{center}
      \end{minipage}
      \hspace{-0.9cm}
      \begin{minipage}{64mm}
        \begin{center}
          \includegraphics[width=64mm]{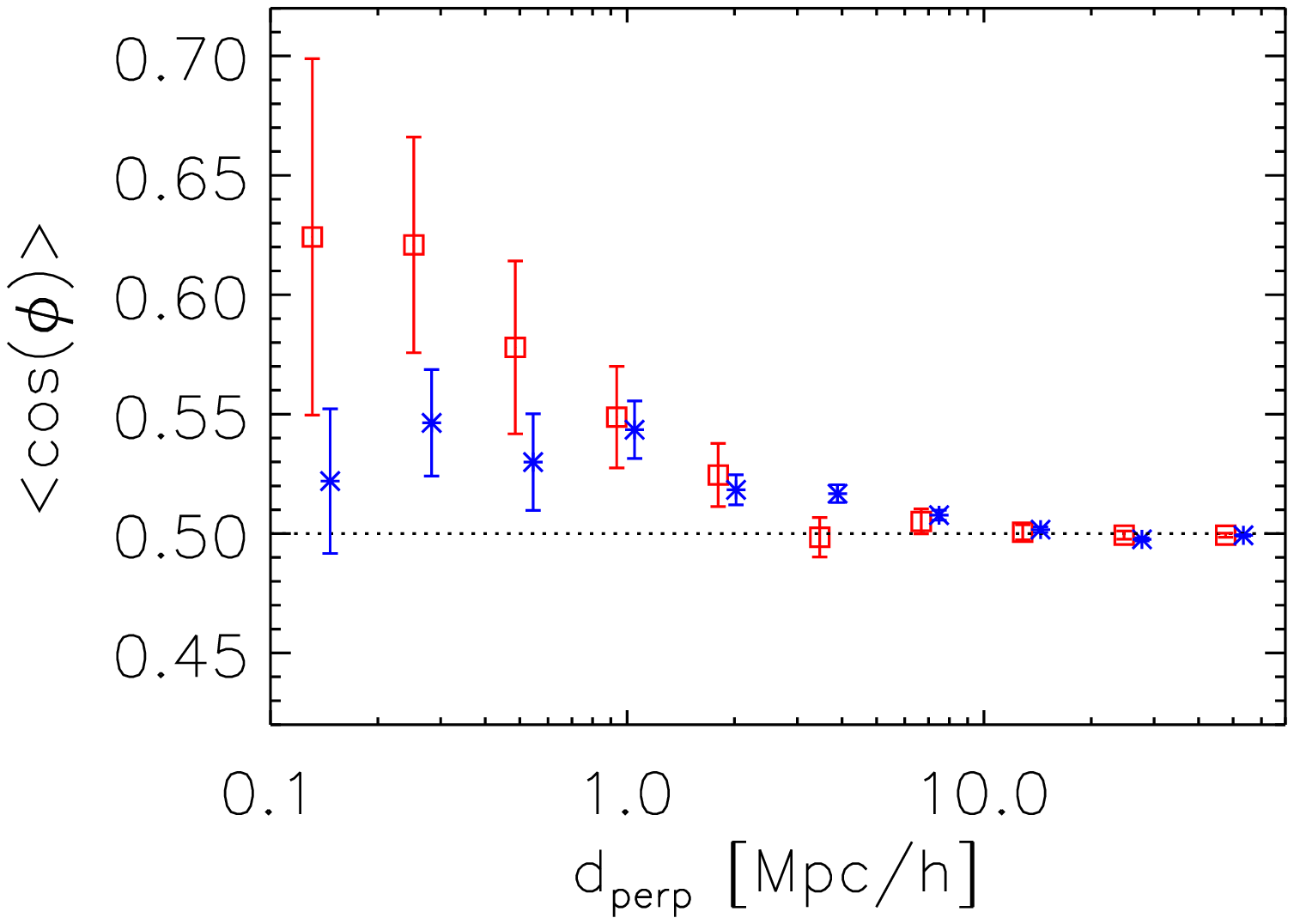}
        \end{center}
      \end{minipage}
      \hspace{-0.9cm}
      \begin{minipage}{64mm}
        \begin{center}
          \includegraphics[width=64mm]{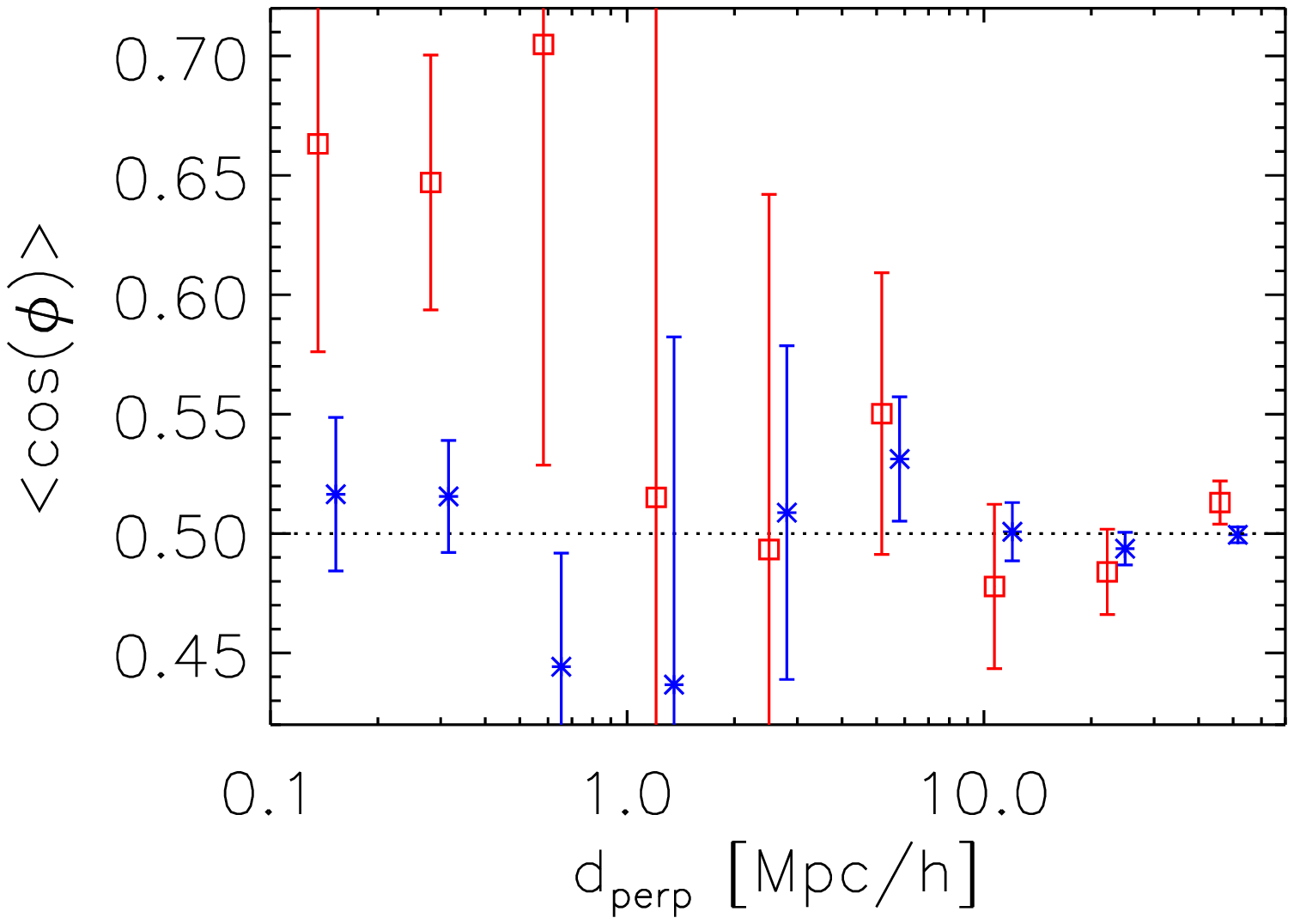}
        \end{center}
      \end{minipage}
    \end{tabular}
    \vspace{-5mm}
    \begin{tabular}{cc}
      \begin{minipage}{64mm}
        \begin{center}
          \includegraphics[width=64mm]{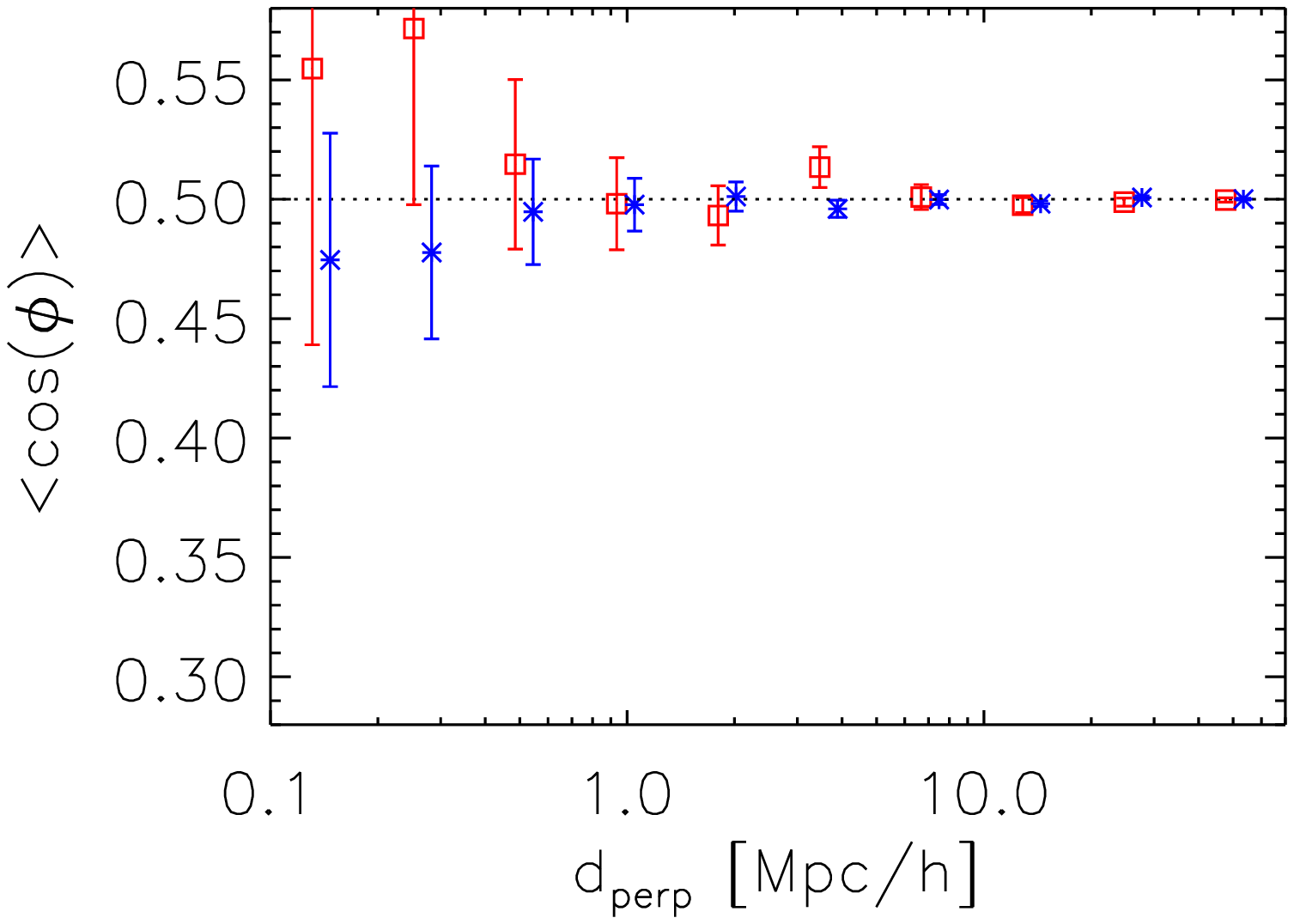}
        \end{center}
      \end{minipage}
      \hspace{-0.9cm}
      \begin{minipage}{64mm}
        \begin{center}
          \includegraphics[width=64mm]{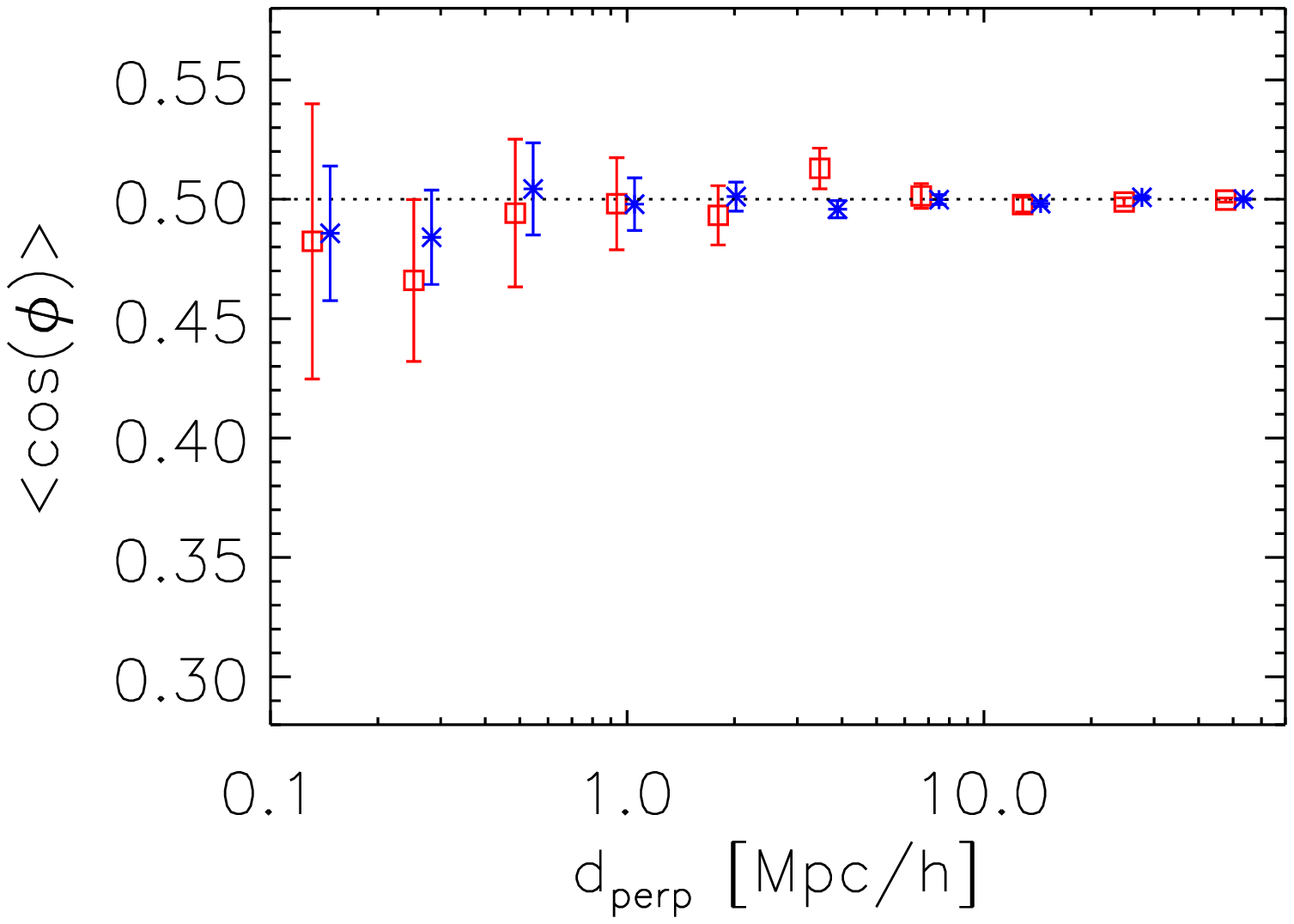}
        \end{center}
      \end{minipage}
      \hspace{-0.9cm}
      \begin{minipage}{64mm}
        \begin{center}
          \includegraphics[width=64mm]{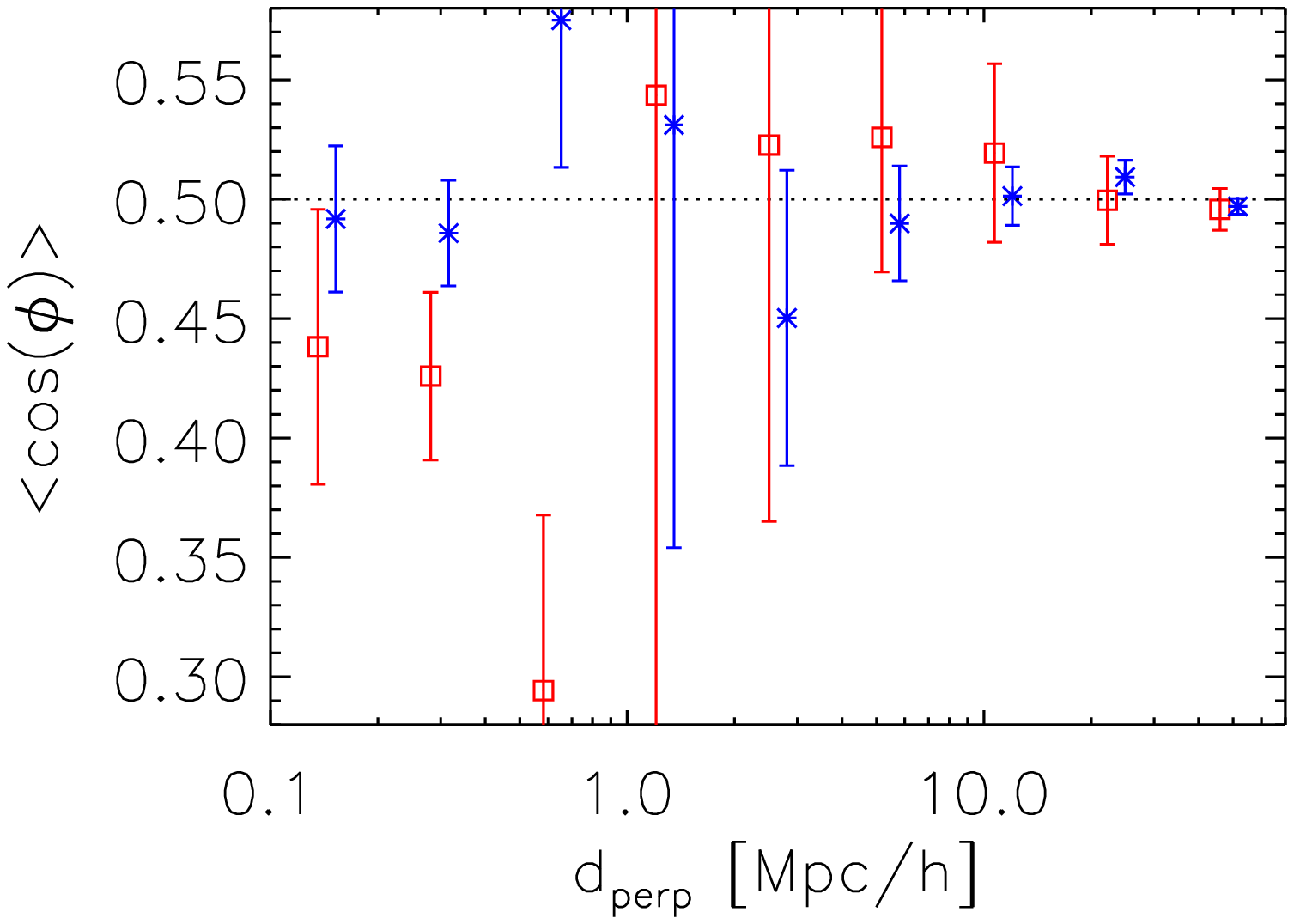}
        \end{center}
      \end{minipage}
    \end{tabular}
    \vspace{-5mm}
    \begin{tabular}{cc}
      \begin{minipage}{64mm}
        \begin{center}
          \includegraphics[width=64mm]{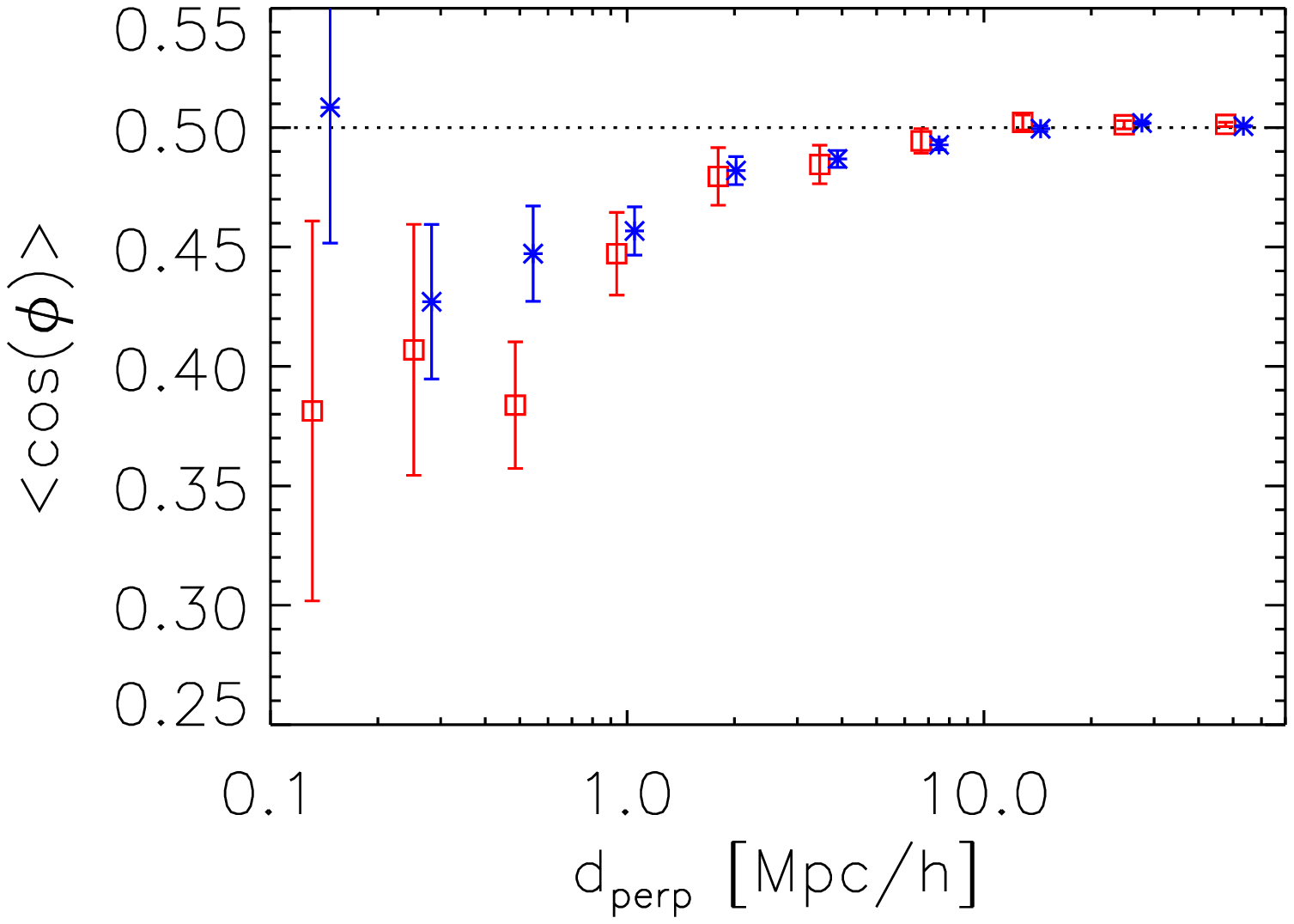}
        \end{center}
      \end{minipage}
      \hspace{-0.9cm}
      \begin{minipage}{64mm}
        \begin{center}
          \includegraphics[width=64mm]{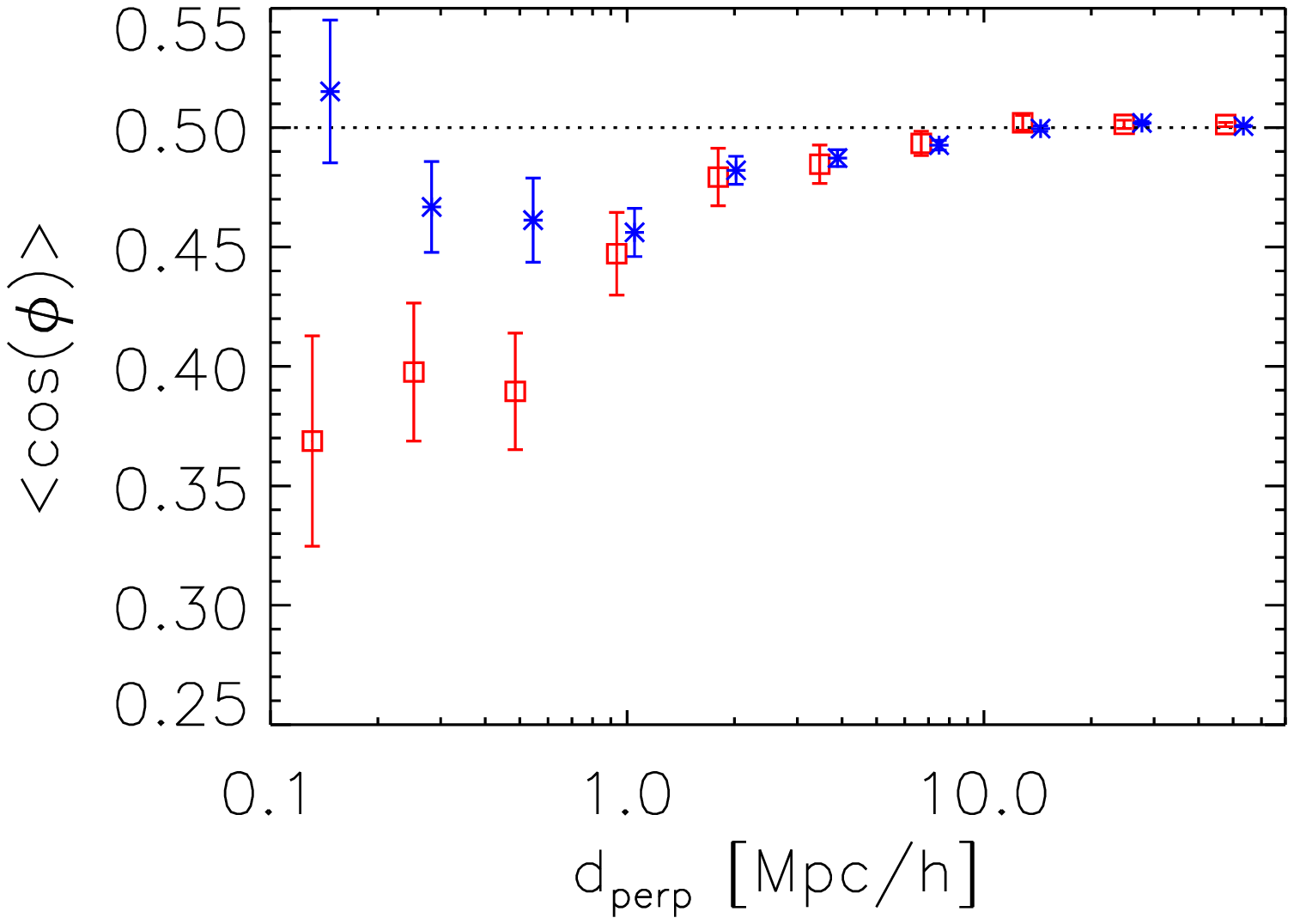}
        \end{center}
      \end{minipage}
      \hspace{-0.9cm}
      \begin{minipage}{64mm}
        \begin{center}
          \includegraphics[width=64mm]{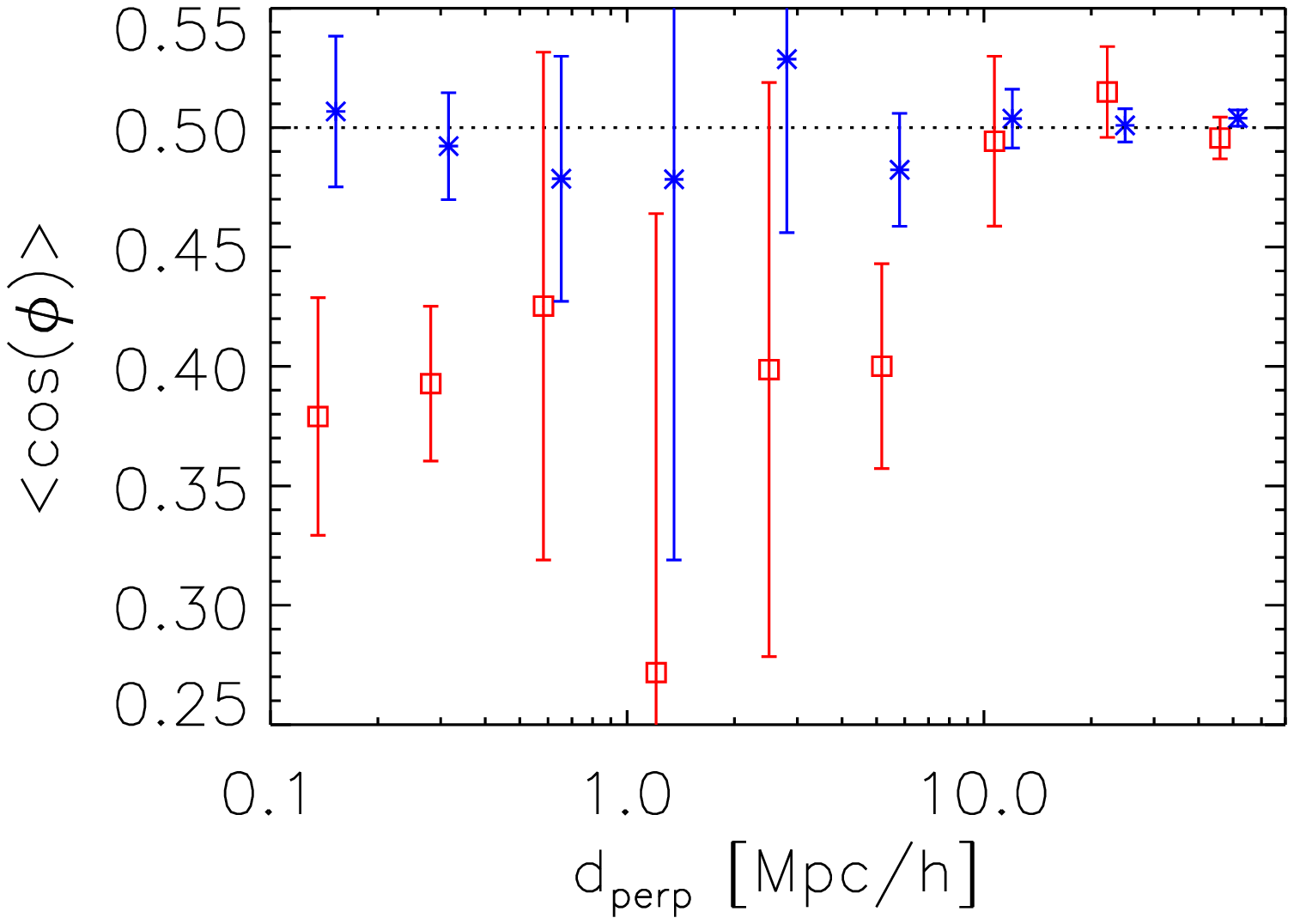}
        \end{center}
      \end{minipage}
    \end{tabular}
  \end{center}
  \caption{Alignments of the principal axes of the halos with the cluster--cluster axis as 
           a function of perpendicular distance $d_{\mbox{perp}}$ from that axis. The rows show 
           the major, intermediate, and minor axes (from top to bottom). In the leftmost, center, 
           and rightmost columns we plot alignments of halos, halos plus clusters, and only clusters,
           respectively. Cluster pairs which are connected by a filament are shown using squares, 
           whereas an asterisk is used for cluster pairs for which the cluster--cluster axis cuts 
           through a void. The dotted line shows the expectation for a random sample with no
           alignments.}
  \label{fig:halo-fil-perpd-align}
\end{figure*}

\section{Alignments of Halos} \label{alignments}

\subsection{Halo Alignments in Filaments}

For each cluster pair, we examine the halos that lie in the cylinder whose central axis is defined by the
cluster--cluster axis and whose radius extends 70\,$h^{-1}$\,Mpc from the cluster--cluster axis. By
going out to these large radii we cover as much of the simulation volume for each cluster 
pair as possible. We define two vectors $\uv{u}_1$ and $\uv{u}_2$ such that $\uv{u}_1$ lies on the
cluster--cluster axis, and $\uv{u}_2$ points along one of the halo principal axes. Our
measure of the alignment between them is 
defined by
\begin{equation}
\vert cos(\phi) \vert\ =\vert \uv{u}_1 \cdot \uv{u}_2 \vert 
\end{equation}
where $\phi$ is the angle between the two vectors.
For each of the three eigenaxes we compute this alignment.

\subsubsection{Filament--Halo Alignments}

In Figure~\ref{fig:halo-fil-perpd-align}, we plot the alignments of the principal axes of the
halos with the cluster--cluster axis as a function of the perpendicular distance from that
axis. The rows show the major, intermediate, and minor axes. In the leftmost,
center, and rightmost columns we plot alignments of halos, of halos plus clusters, and of clusters,
respectively. Cluster pairs which are connected by a filament are shown using squares, whereas
asterisks are used for cluster pairs for which no coherent structure was
found along the cluster--cluster axis.
Errors bars are computed assuming Poissonian statistics. The dotted line shows the expectation for 
a random sample with no alignments.

For small separations from the cluster--cluster axis, the major and minor axes of halos are 
aligned and anti--aligned with that axis, respectively (leftmost column of 
Figure~\ref{fig:halo-fil-perpd-align}). Interestingly, there is no clear difference
between those cluster pairs that are connected by a filament and those that are not connected
at all. For galaxy--size halos therefore the fact that they lie in a filament or not does not affect
their alignments (we will examine the effect on their ellipticities in Section 4.2).

The alignment signals become small at larger separations from the 
cluster--cluster axis.  At separations of around 4\,$h^{-1}$\,Mpc the alignment signal is 
almost absent. As can be seen from Figure~8 in Colberg et al. (2005), at this scale,
the averaged density profile of filaments has dropped strongly from its central value.
However, given
 that we do not find a difference in the alignments between connected and unconnected
cluster pairs, this finding has to be treated as a mere coincidence.

We note that the (anti) alignment signal is somewhat stronger for the minor axes than for
major axes. Also, for the minor
axes, there is a small difference between halos in filaments and halos elsewhere, although the
error bars are relatively large. The intermediate halo axes are neither aligned nor anti--aligned
with the cluster--cluster axes.

\subsubsection{Filament--Cluster Alignments}

The rightmost column of Figure~\ref{fig:halo-fil-perpd-align} shows the alignment results
for clusters. The major and minor axes of the clusters connected by filaments are aligned
and anti--aligned with those filaments, respectively. 
Although the statistical uncertainty is quite large, the signal is significantly stronger than for
galaxy--size halos.
However, no such signal exists for cluster 
pairs that are not connected by filaments, which again
is very different from the galaxy halo result. This finding provides an explanation for the Bingelli 
effect, which is directly connected to the presence of filaments.

\begin{figure*}
  \begin{center}
    \begin{tabular}{cc}
      \begin{minipage}{64mm}
        \begin{center}
          \includegraphics[width=64mm]{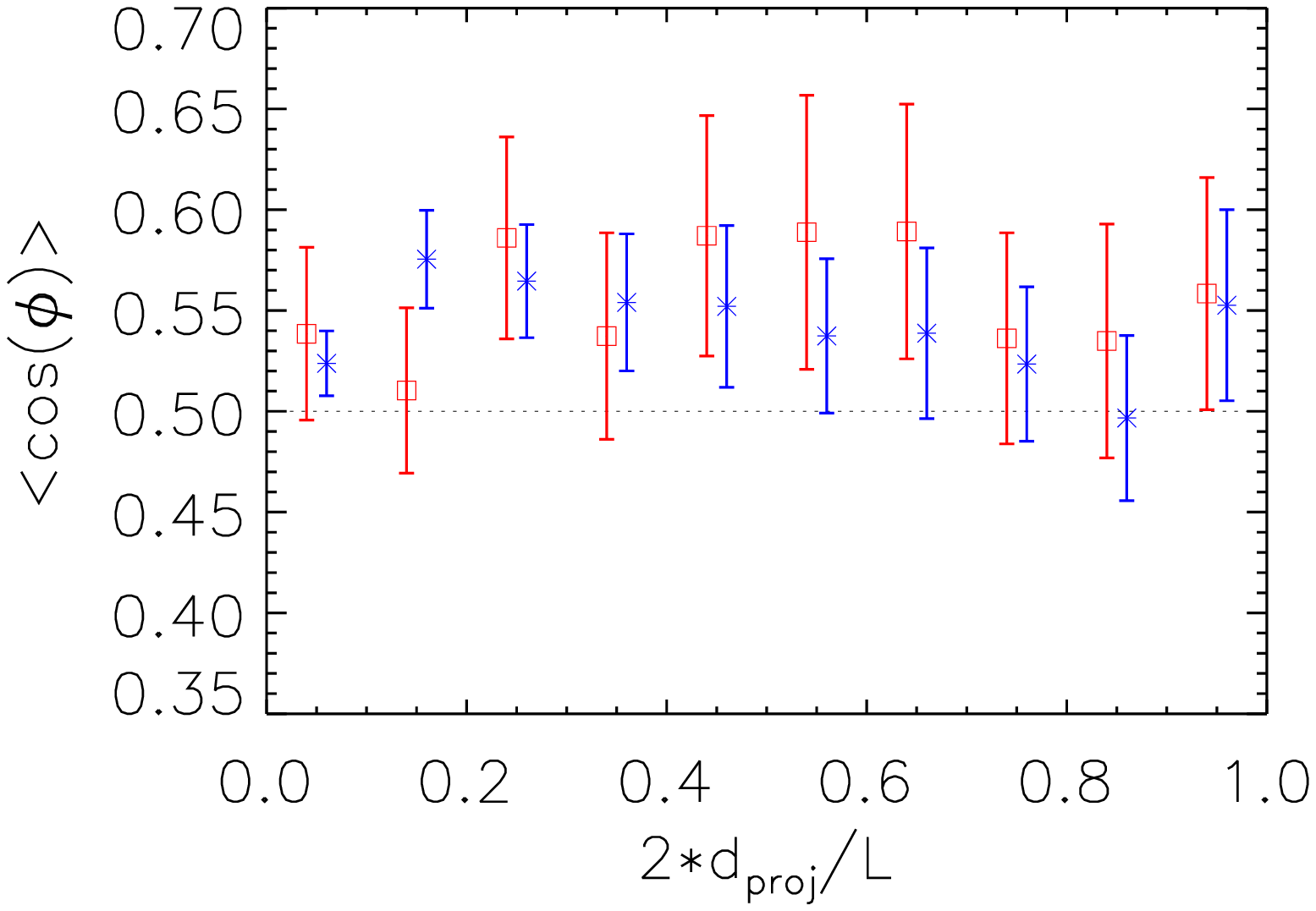}
        \end{center}
      \end{minipage}
      \hspace{-0.9cm}
      \begin{minipage}{64mm}
        \begin{center}
          \includegraphics[width=64mm]{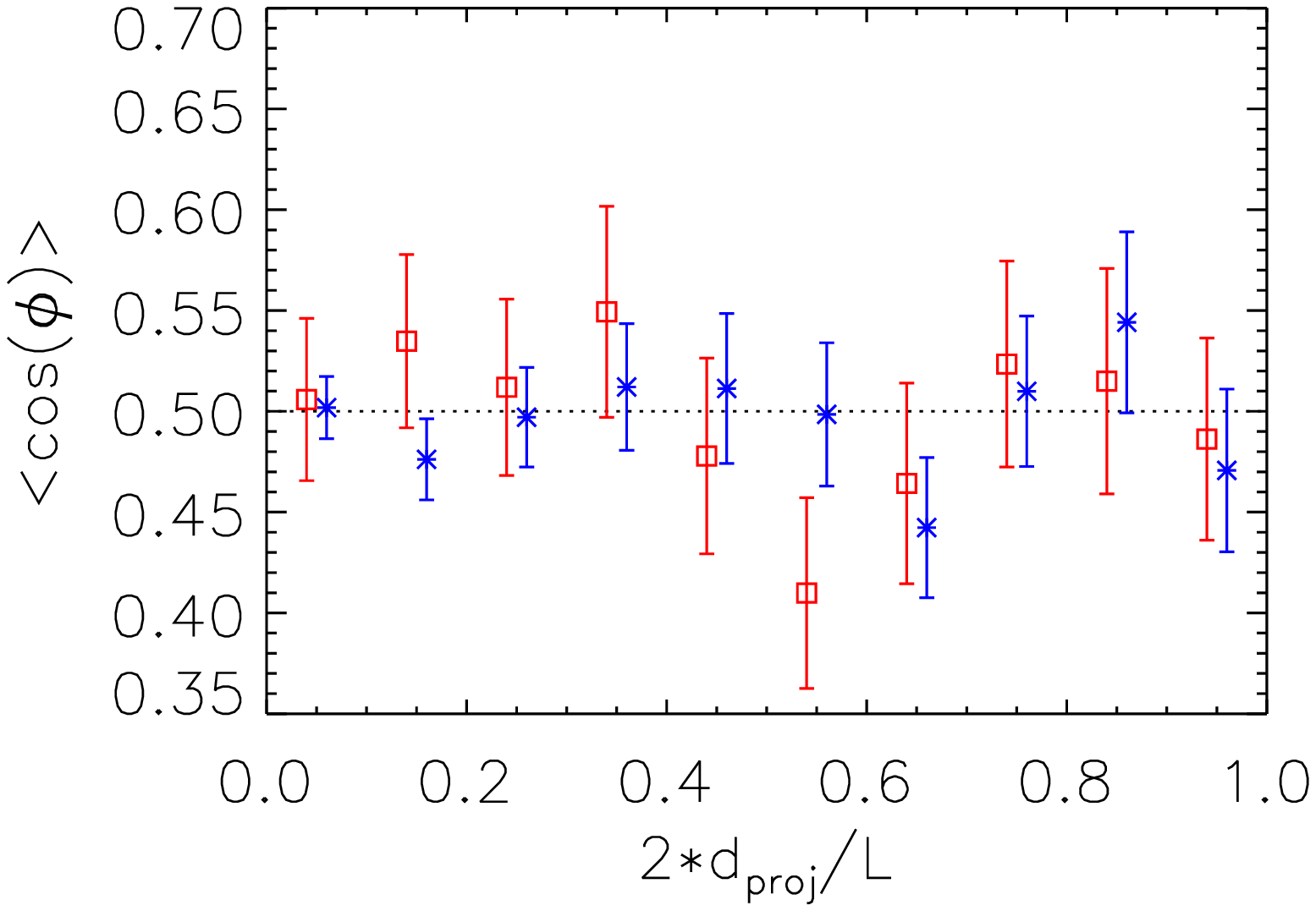}
        \end{center}
      \end{minipage}
      \hspace{-0.9cm}
      \begin{minipage}{64mm}
        \begin{center}
          \includegraphics[width=64mm]{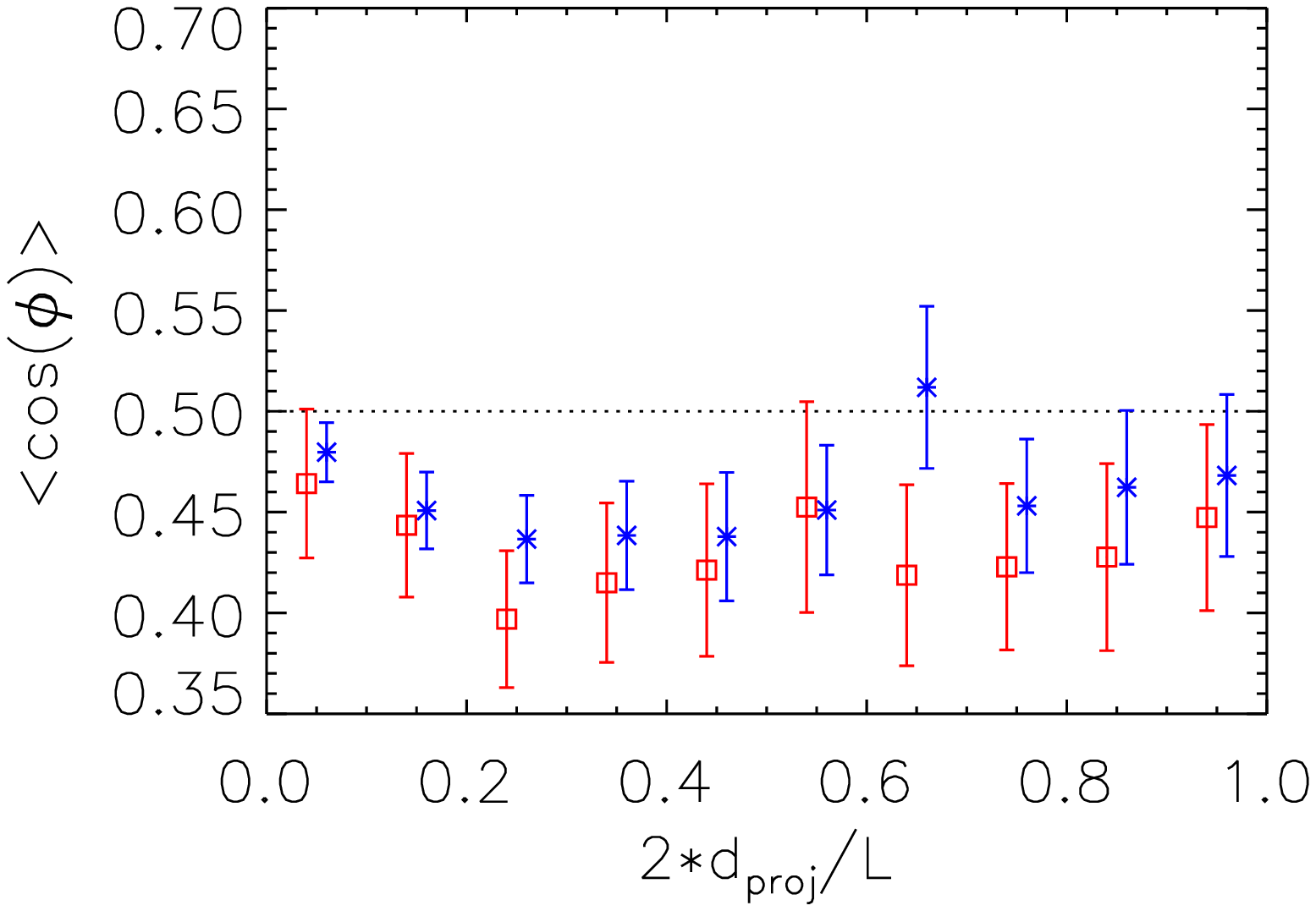}
        \end{center}
      \end{minipage}
    \end{tabular}
    \vspace{-5mm}
    \begin{tabular}{cc}
      \begin{minipage}{64mm}
        \begin{center}
          \includegraphics[width=64mm]{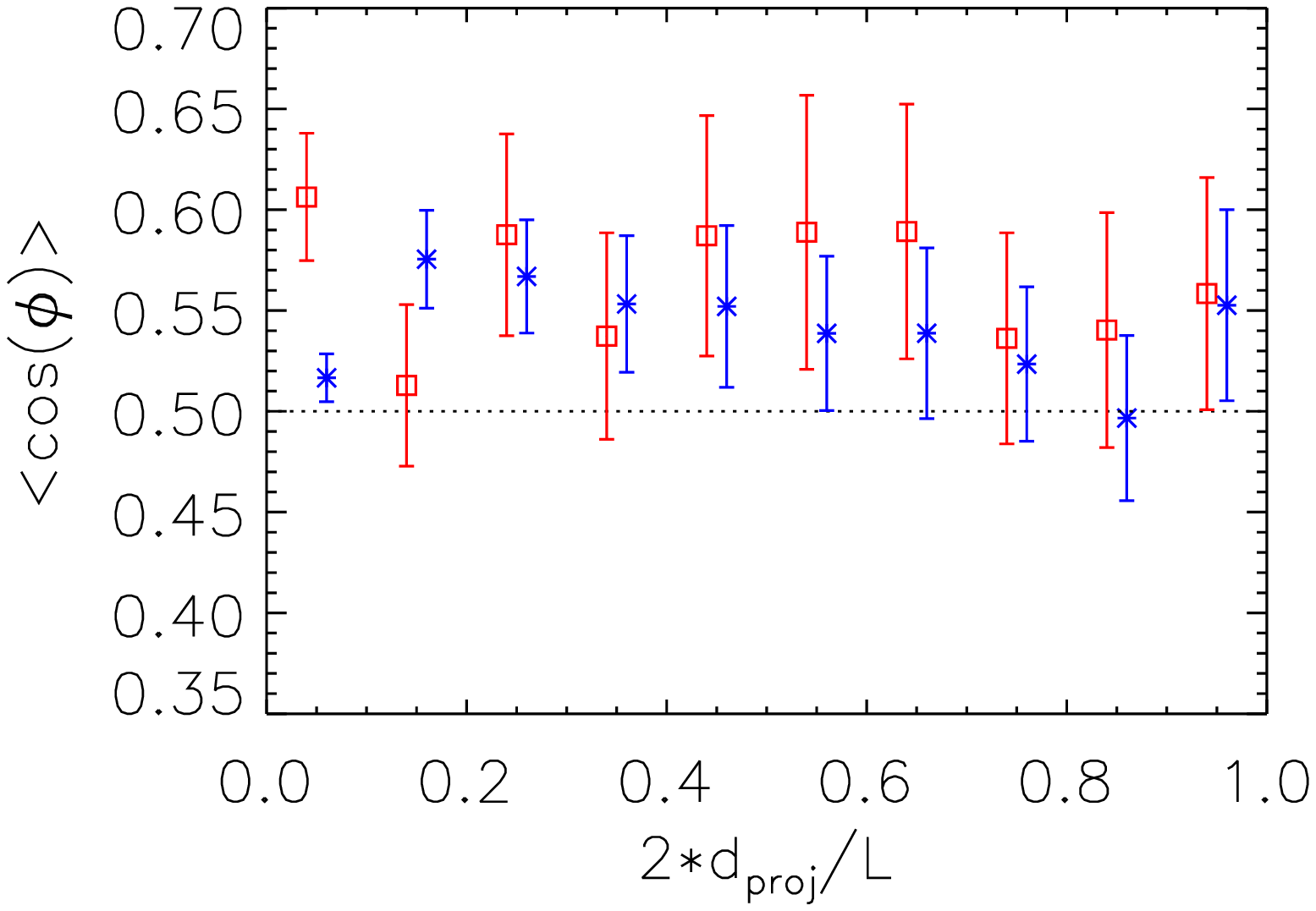}
        \end{center}
      \end{minipage}
      \hspace{-0.9cm}
      \begin{minipage}{64mm}
        \begin{center}
          \includegraphics[width=64mm]{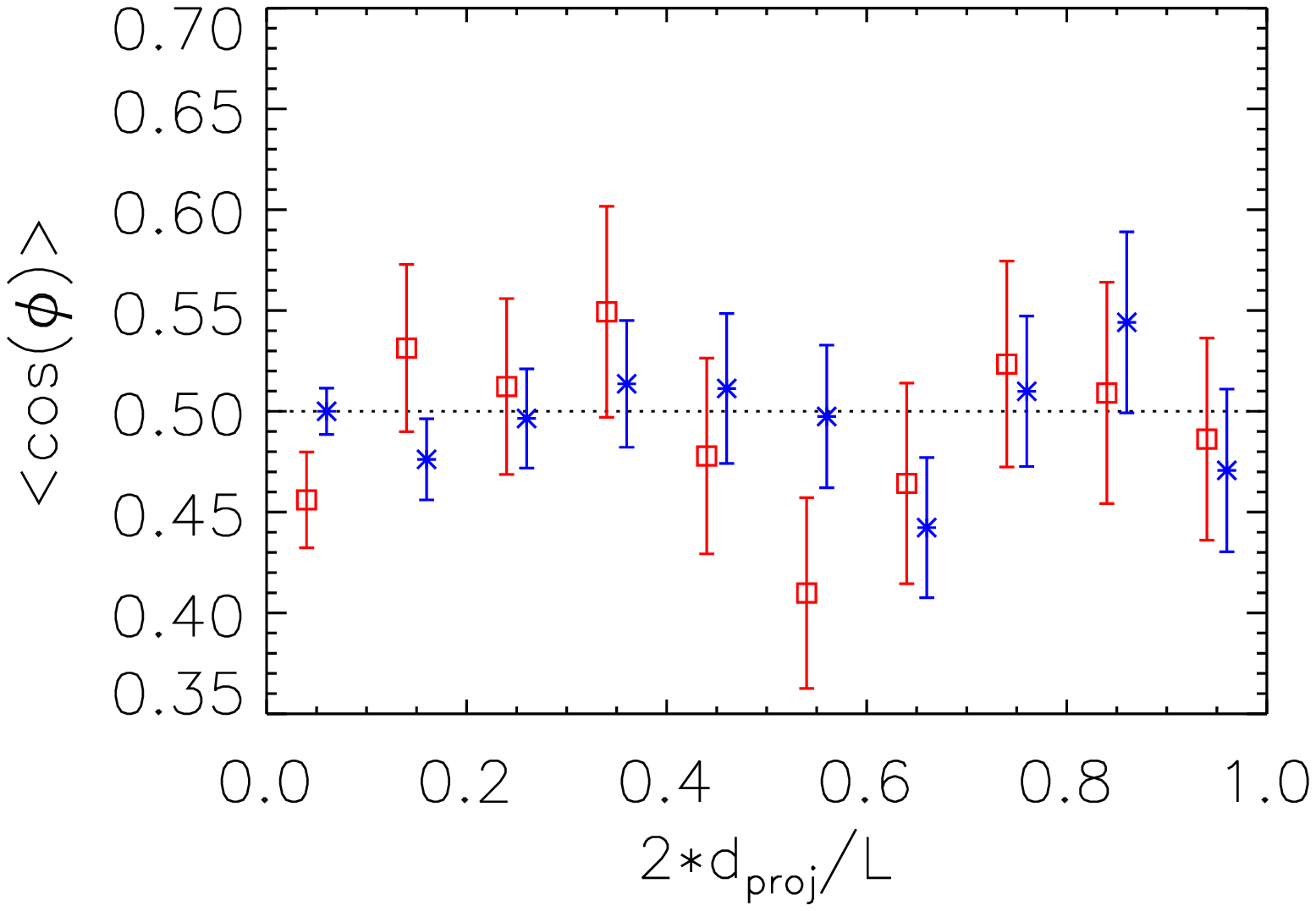}
        \end{center}
      \end{minipage}
      \hspace{-0.9cm}
      \begin{minipage}{64mm}
        \begin{center}
          \includegraphics[width=64mm]{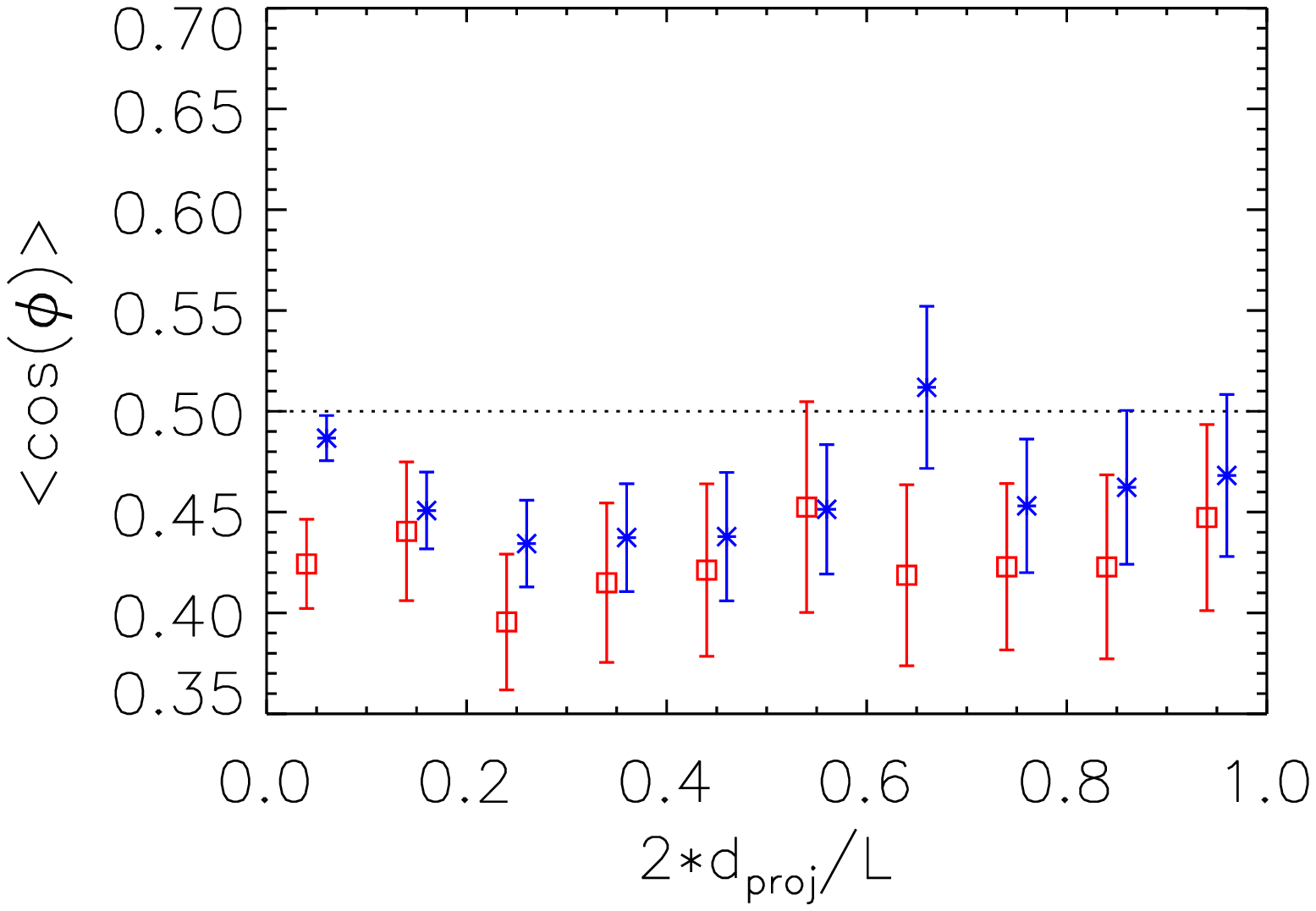}
        \end{center}
      \end{minipage}
    \end{tabular}
   \end{center}
  \caption{Alignments for the major (leftmost column), intermediate (middle column), and 
           minor (rightmost column) halo axes, both for halos (top row) and for halos 
           plus clusters (bottom row). The symbols are the same as in 
           Figure~\ref{fig:halo-fil-perpd-align}. See main body for an explanation of
           the horizontal axis of each plot. The dotted line shows the expectation for a 
           random sample with no alignments.}
  \label{fig:halo-fil-projd-align}
\end{figure*}

\subsection{Alignments as a Function of Projected Distance from the Cluster--Cluster Axis}

Having examined alignments as a function of the radial distance from cluster--cluster axes,
we now turn our attention to the dependence of any alignment signal on the
distance of a halo from its nearest cluster, i.e. along the cluster--cluster axis. We
restrict this analysis to all halos that are within 1.5\,$h^{-1}$\,Mpc from the axis.

Figure~\ref{fig:halo-fil-projd-align} shows the alignments for the
three halo axes, both for halos (top
row) and for halos plus clusters (bottom row). The symbols are the same as in 
Figure~\ref{fig:halo-fil-perpd-align}. We scale the halo--cluster separations
on the $x$--axis such that
a halo position just outside the virial radius of the  nearest cluster is at position 0, whereas a halo right in the center between
both clusters sits is at position 1. $L$ is the total length of a cluster--cluster axis, and $d_{\mbox{proj}}$ 
corresponds to the distance of a halo from the nearer one of the clusters that define
the cluster--cluster axis. As before, the dotted line shows the expectation for a random 
sample with no alignments.

Figure~\ref{fig:halo-fil-projd-align} shows that the magnitude of the alignments does not 
depend on the distance of a halo from the nearest cluster. There also is no difference
between cluster pairs that are connected by a filament and those that intersect a void.

\begin{figure*}
  \begin{center}
    \begin{tabular}{cc}
      \begin{minipage}{63mm}
        \begin{center}
          \includegraphics[width=63mm]{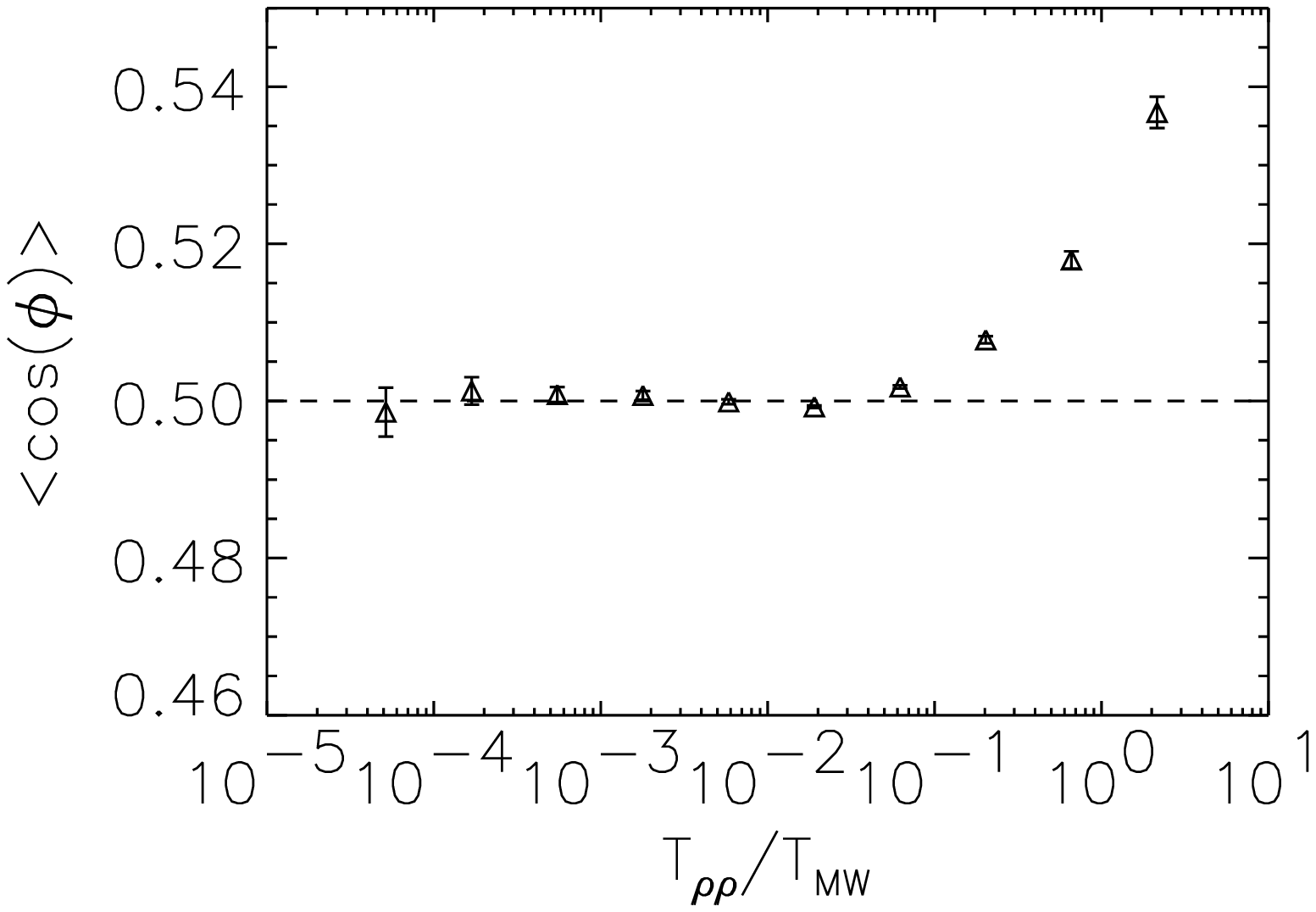}
        \end{center}
      \end{minipage}
      \hspace{-0.7cm}
      \begin{minipage}{63mm}
        \begin{center}
          \includegraphics[width=63mm]{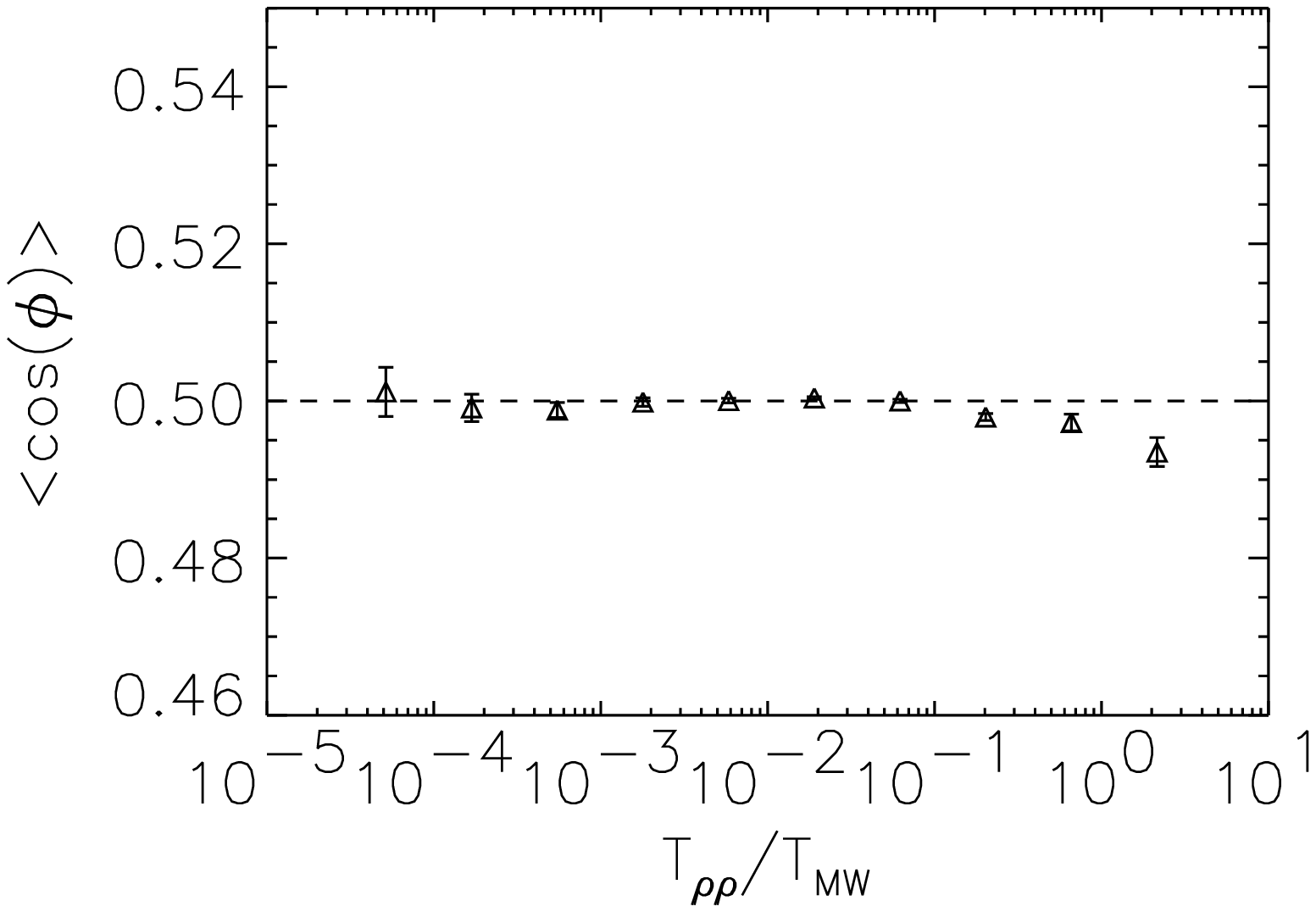}
        \end{center}
      \end{minipage}
      \hspace{-0.7cm}
      \begin{minipage}{63mm}
        \begin{center}
          \includegraphics[width=63mm]{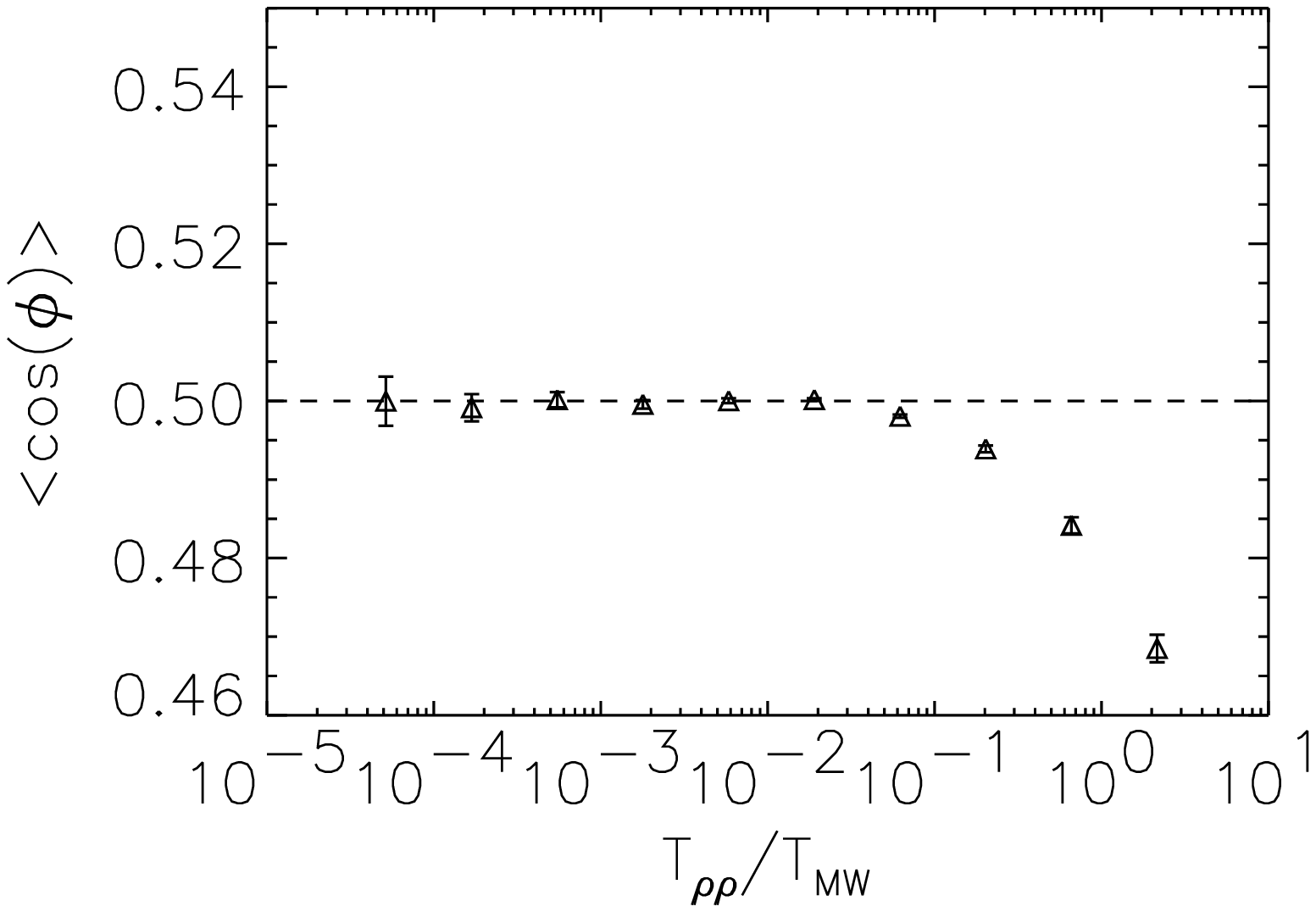}
        \end{center}
      \end{minipage}
    \end{tabular}
    \vspace{-5mm}
    \begin{tabular}{cc}
      \begin{minipage}{63mm}
        \begin{center}
          \includegraphics[width=63mm]{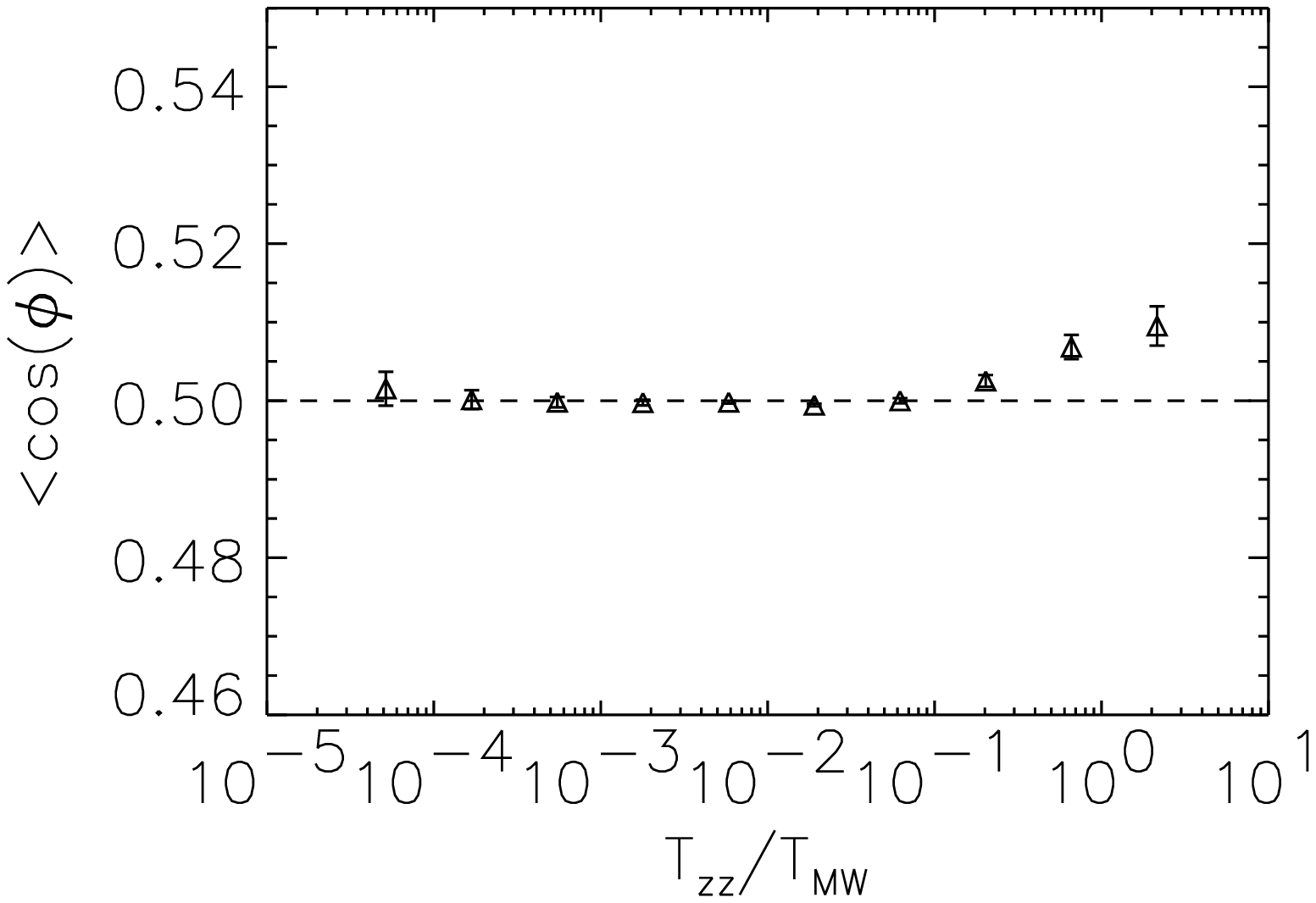}
        \end{center}
      \end{minipage}
      \hspace{-0.7cm}
      \begin{minipage}{63mm}
        \begin{center}
          \includegraphics[width=63mm]{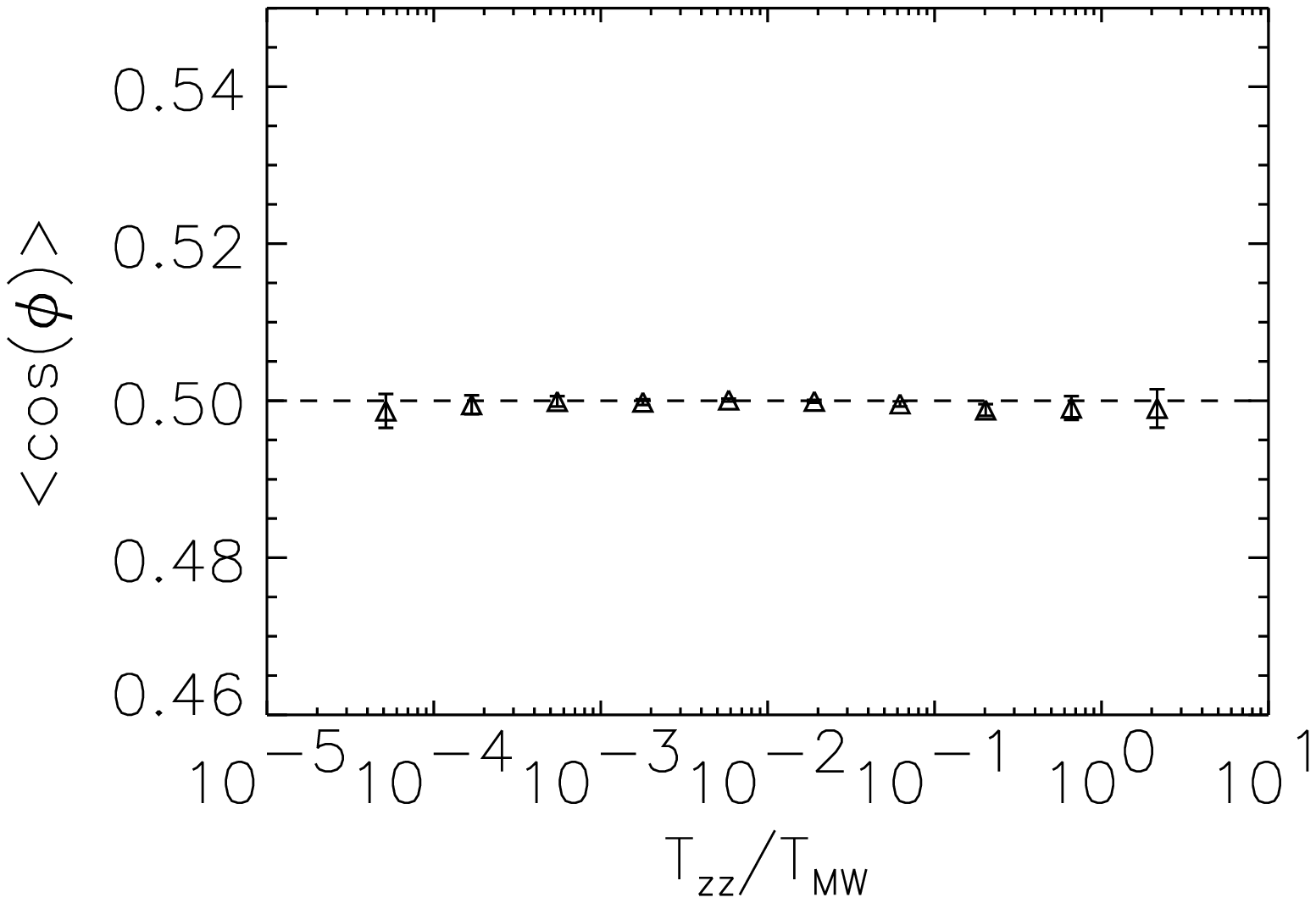}
        \end{center}
      \end{minipage}
      \hspace{-0.7cm}
      \begin{minipage}{63mm}
        \begin{center}
          \includegraphics[width=63mm]{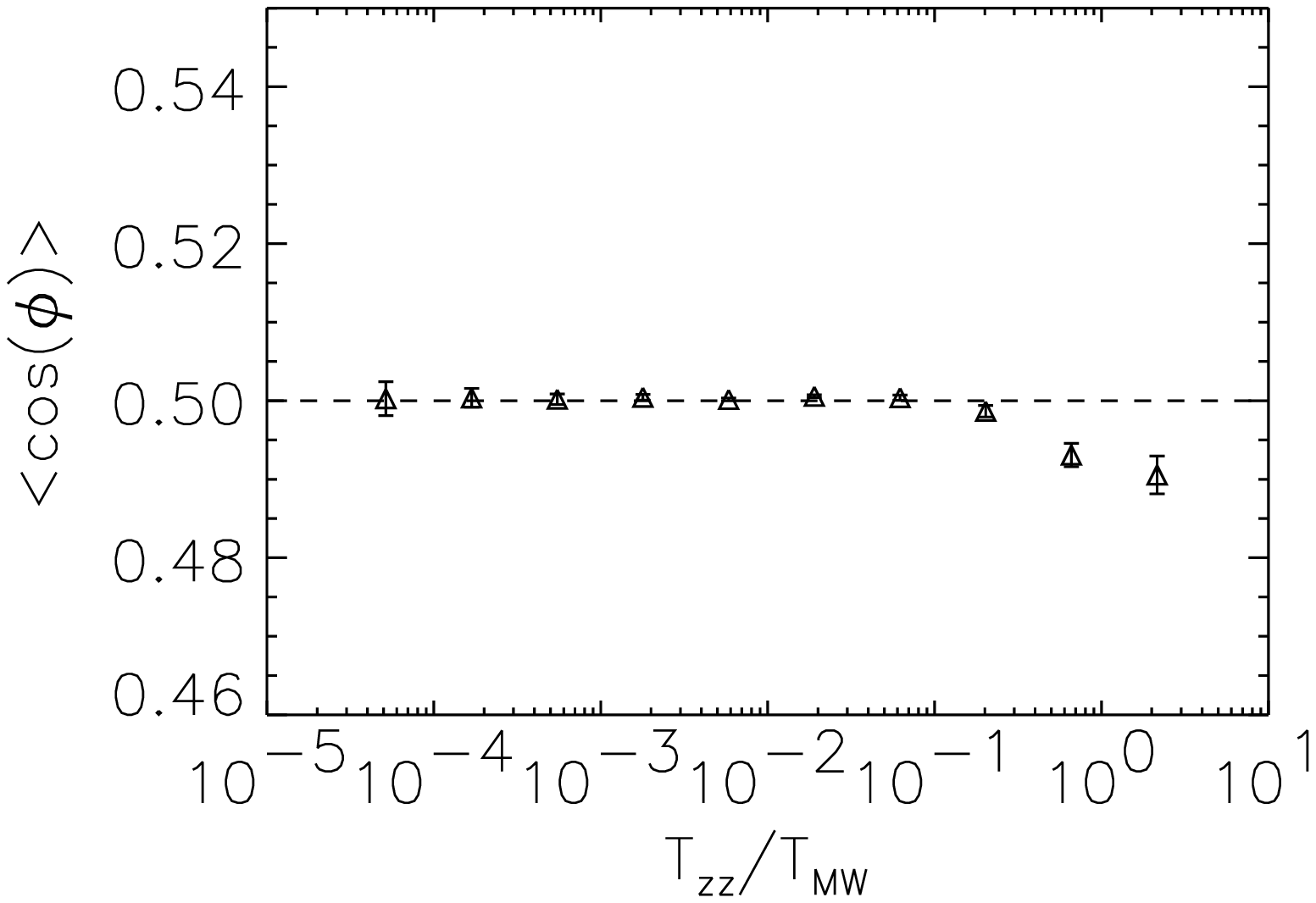}
        \end{center}
      \end{minipage}
    \end{tabular}
   \end{center}
  \caption{Alignments for the major (leftmost column), intermediate (middle column), and 
           minor (rightmost column) halo axes as a function of local tidal fields, with stretching 
           perpendicular to the cluster--cluster axis (top row), and parallel to the 
           cluster--cluster axis (bottom row).  The dotted line shows the expectation for 
           a random sample with no alignments. The suffix "`MW"' refers to the re--scaling
           of the tidal field using that of Milky Way sized halo at distance of  1 h$^{-1}$ Mpc.}
  \label{fig:tidal-align}
\end{figure*}

\subsection{Alignments as a Function of Cluster--Cluster Tidal Fields}

The presence of tidal fields can provide a very natural explanation for the alignments  
and shapes of objects. In the following sections, we 
will study the influence of the tidal forces exterted by the clusters that are used to 
define the cluster--cluster configurations.

\subsubsection{Modeling the Clusters}

In order to make a simple model of the tidal field, we treat the two clusters as spherical objects
 and we ignore the mass contained
in the filaments. Given the large masses of the clusters and the relatively
low average overdensity of filaments (see the discussion in Colberg et al. 2005) this simple
model is reasonable. 

For this dumbell--shaped configuration, use of an analytical expression for the
gravitational potential is possible. We use cylindrical coordinates ($\rho, \theta, z$) and take
one of the clusters to lie at the origin with the other one on the $z$--axis. The
gravitational potential $\Phi$ then becomes
\begin{equation}
\Phi=\frac{M_{1}}{\sqrt{\rho^2+z^2}} +
\frac{M_{2}}{\sqrt{\rho^2+(z-L)^2}}\,,
\end{equation} 
where $L$ is the the length of the cluster--cluster axis, and the masses of the
two clusters are given by $M_1$ and $M_2$. The components of the tidal field tensor $T$
can be computed through 
\begin{equation}
T_{ij} = \frac{\partial^2 \Phi}{\partial x_{i} \partial x_{j}}\,.
\end{equation}
Since we are only interested in the absolute magnitude of the tidal field components we 
neglect the sign, thus taking the absolute value to be a measure of the stretching a body 
would feel.  

In this axially symmetric model, there are two components of $T_{ij}$ that stretch a 
spherical mass into an ellipse. They are
\begin{eqnarray}
T_{zz} & = &
\frac {M_2 [2(L-z)^2-\rho^2]} {[\rho^2+(L-z)^2]^{5/2}} 
- \frac {M_1 (\rho^2-2z^2)} {(\rho^2+z^2)^{5/2}} \\
\mbox{and}\nonumber\\
T_{\rho\rho} & = & 
\frac {M_1 (2\rho^2-z^2)} {(\rho^2+z^2)^{5/2}}
- \frac{M_2 [(L-z)^2-2\rho^2]}{[\rho^2+(L-z)^2]^{5/2}}      
\end{eqnarray}
$T_{zz}$ stretches along the cluster--cluster axis, whereas $T_{\rho\rho}$ stretches 
perpendicular to the cluster--cluster axis.

We normalize the tidal field components by dividing them by the strength of the tidal
field of a Milky--Way size halo with $M_{\mbox{MW}} = 2.5\cdot\,10^{12}\,h^{-1}\,M_\odot$
at a radius of $1\,h^{-1}$\,Mpc. We use a procedure similar to the one previously outlined 
to compute the alignments, as follows. 
To study $T_{\rho\rho}$, we take $\uv{u}_1$ to be perpendicular to the cluster--cluster 
axis and pointing towards the halo and $\uv{u}_2$ along one of the principal axes of the 
halo as before. 

\subsubsection{Alignments in the Tidal Field}

Figure~\ref{fig:tidal-align} shows the alignments for the two tidal--field
components $T_{zz}$ and $T_{\rho\rho}$. As before, the leftmost, center, and
rightmost columns show the major, intermediate, and minor halo axes, respectively.
The two rows give the alignments along $T_{zz}$ and $T_{\rho\rho}$. In each plot,
the dotted line shows the expectation for a random sample with no alignments.

The stretching perpendicular to the cluster--cluster axis is about twice that  
parallel to the cluster--cluster axis.  We note that for this analysis we do not
differentiate between cluster pairs with filaments or voids in between. This
also results in smaller error bars, since the sample sizes are bigger. The 
magnitude of this effect is smaller than alignments discussed in the previous 
sections. However, unlike the alignment as a function of distance from the
cluster-cluster axis, the alignment with the tidal field is a much more
abrupt function of tidal field strength. From 
Figure~\ref{fig:tidal-align} we can see that when the tidal field is less than $\sim 10 \%$
of the MW  field the results are statistically consistent with no alignment.
The sudden change for larger values of the tidal field indicates the direct role
that tidal distortions must play in aligning halos.

\begin{figure}
  \includegraphics[width=85mm]{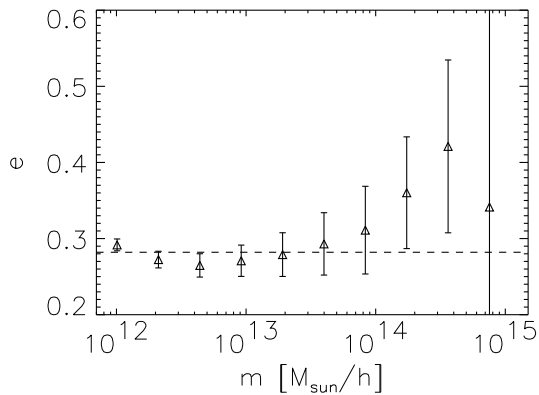}
  \caption{Halo and cluster ellipticities as a function of mass. The dotted line is the random 
           expectation of the ellipticity.}
  \label{fig:ellip-mass}
\end{figure}

\begin{figure*}
  \begin{center}
    \begin{tabular}{cc}
      \begin{minipage}{64mm}
        \begin{center}
          \includegraphics[width=64mm]{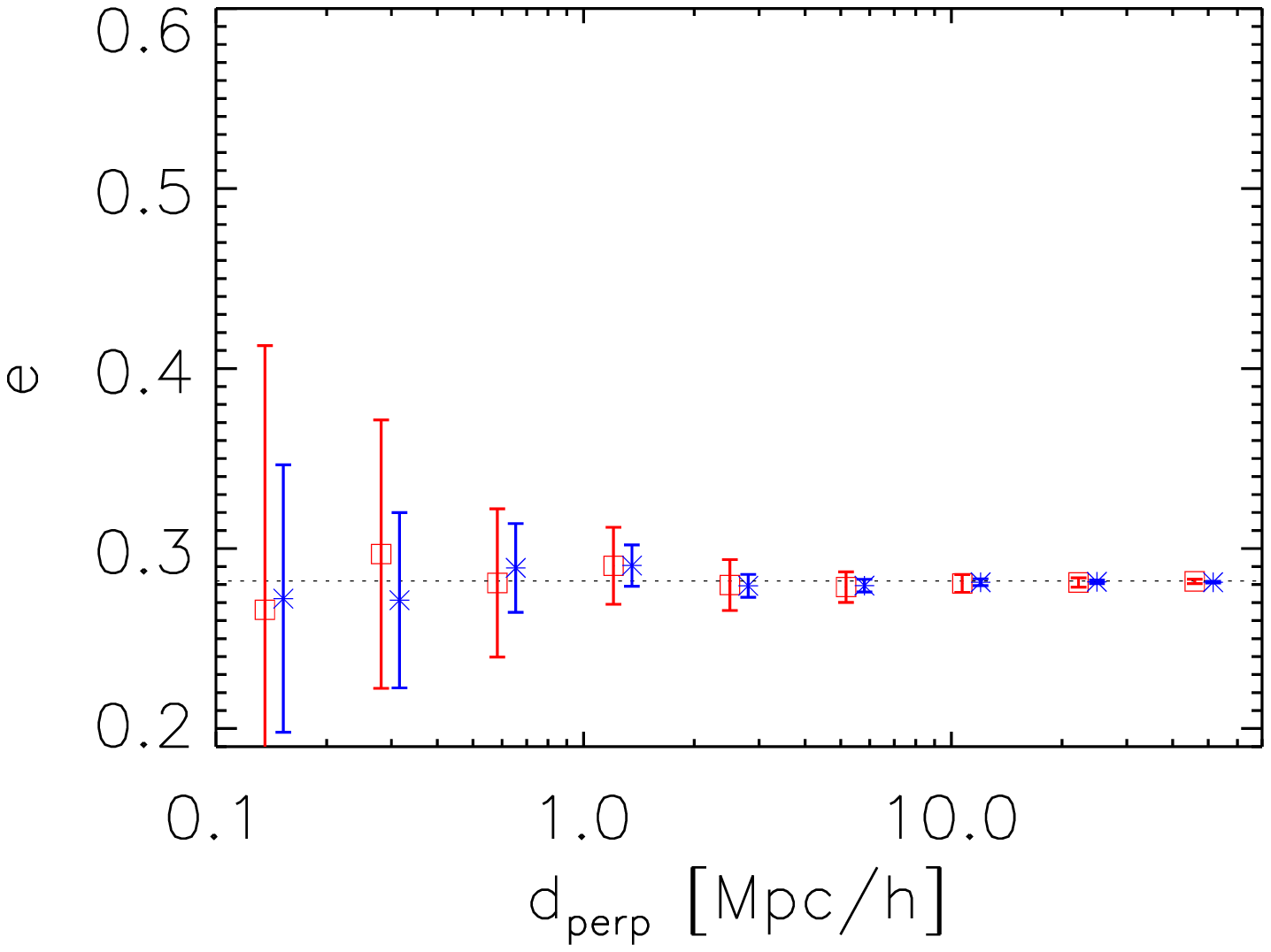}
        \end{center}
      \end{minipage}
      \hspace{-0.9cm}
      \begin{minipage}{64mm}
        \begin{center}
          \includegraphics[width=64mm]{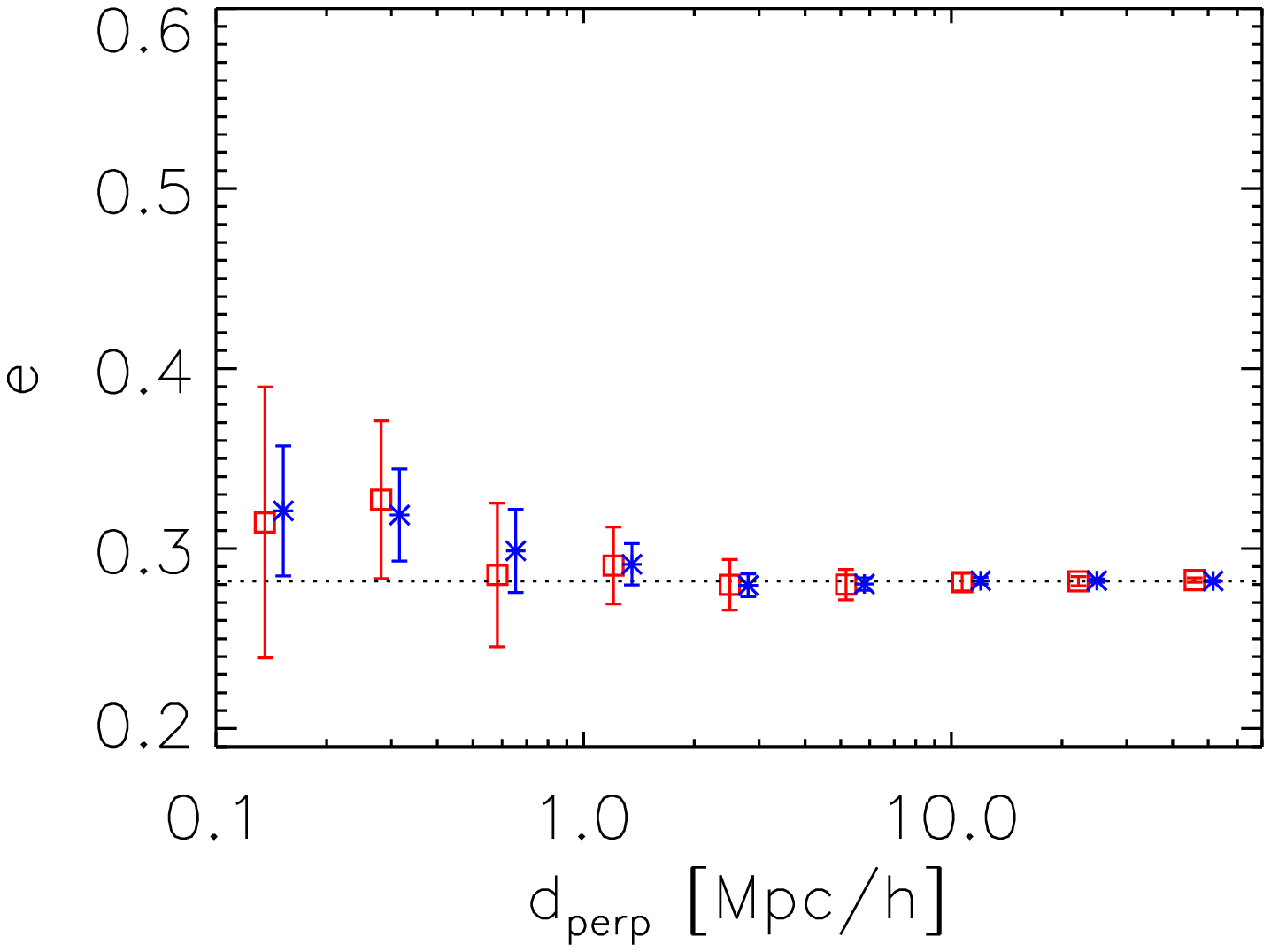}
        \end{center}
      \end{minipage}
      \hspace{-0.9cm}
      \begin{minipage}{64mm}
        \begin{center}
          \includegraphics[width=64mm]{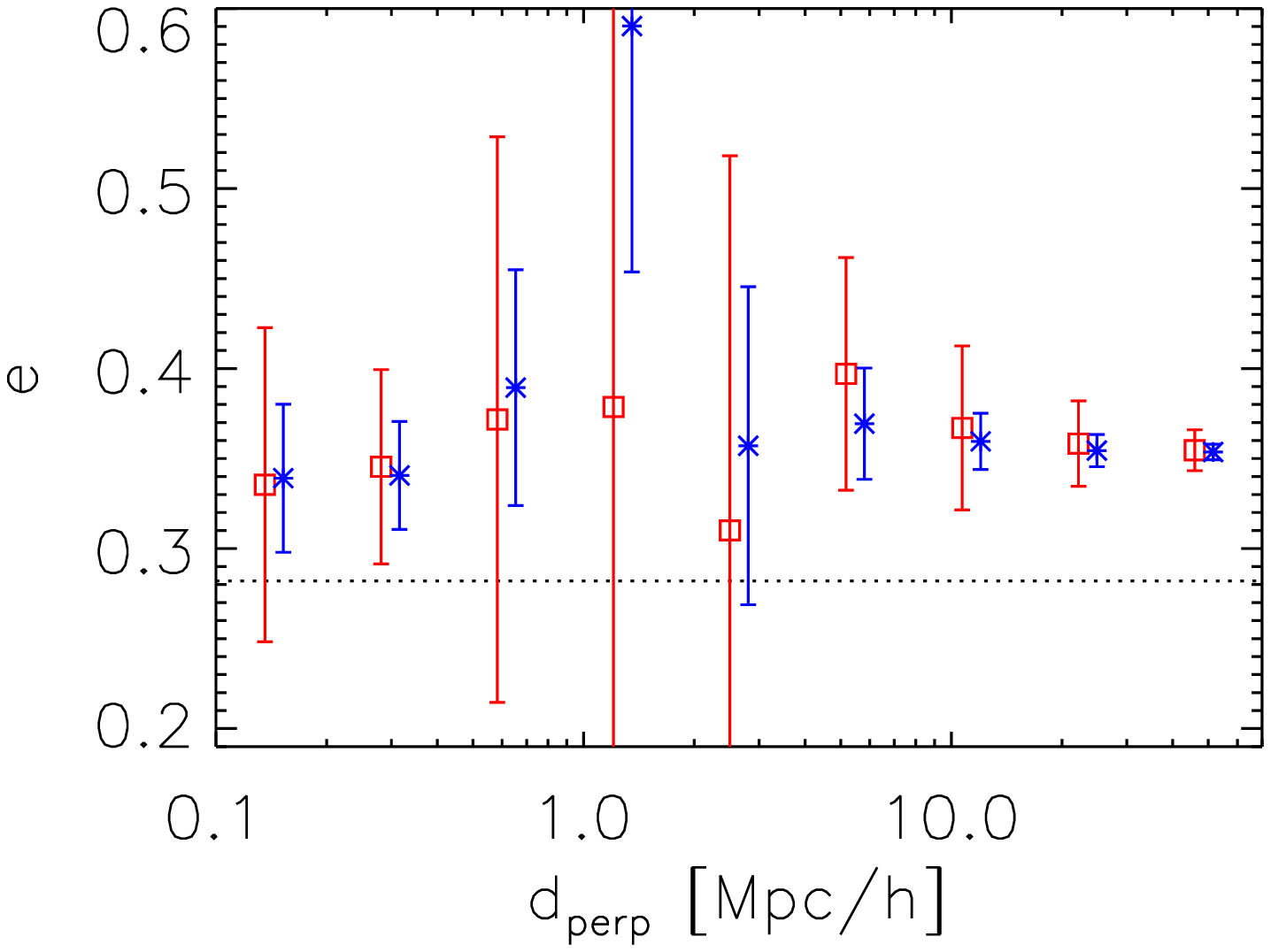}
        \end{center}
      \end{minipage}
    \end{tabular}
  \end{center}
  \caption{Halo and cluster ellipticities as a function of the distance from cluster--cluster
           axes. Shown are all halos (left column), all halos and clusters (middle column), 
           and only the 170 clusters (right column). Square and asterisk symbols denote groups 
           associated with cluster pairs conncected by a filament or with a void in between
           them, respectively. The dotted line is the random expectation.}
  \label{fig:ellip-perpd}
\end{figure*}

\section{Halo and Cluster Ellipticities} \label{ellip}

In the light of the results obtained so far, in particular the alignment signals
and their dependence on the distance from cluster--cluster axes, it is
worthile to examine ellipticities of the halos. 

\subsection{Ellipticity as a Function of Mass}

Figure~\ref{fig:ellip-mass} shows the ellipticities of the halos and clusters in
our sample as a function of their mass. The error bars assume Poissonian distributions.
There is a trend for more massive clusters to be more elliptical than smaller
halos, although the sample size is small.
 This finding agrees with the results obtained by e.g. Warren et al. 1992.

\subsection{Ellipticity as a Function of Distance from Cluster--Cluster Axes}

Figure~\ref{fig:halo-fil-perpd-align} shows that the alignment of halos with a
cluster--cluster axis depends on the distance from that axis. It is thus interesting to
see whether the ellipticities are correlated with that distance as well.

Figure~\ref{fig:ellip-perpd} shows halo and cluster ellipticities as a function of the 
distance from cluster--cluster axes. Shown are all halos (left column), all halos
plus the clusters (center column),  and only the 170 clusters (right column). Square 
and asterisk symbols denote groups associated with cluster pairs conncected by a filament 
or with a void in between them, respectively.

From the figure we see that within the errors
there is no correlation between the ellipticities and the distance of the galaxy halos from 
from the cluster--cluster axes. If tidal forces are responsible for the alignments, they do
not cause a difference in the overall shape of halos.

Furthermore, we also find that there is no difference between halos lying along the 
axes of cluster
pairs connected by a filament and those cluster pairs which have a void in between them.

\section{Summary and Discussion} \label{summary}

Many studies of the alignments of galaxy clusters have been carried out both
with observational data  (Binggeli 
1982, Struble \& Peebles 1985, Flin 1987, Rhee \& Katgert 1987, Ulmer et al. 1989, West 1989, 
Rhee et al. 1992, Plionis 1994, West et al. 1995, Chambers et al. 2000 and 2002) and
theoretically (Splinter et al. 1997, Onuora \& Thomas 2000, Faltenbacher et al. 2002, Kasun \& 
Evrard 2004, Hopkins et al. 2005, Basilakos et al. 2005). Most observations and all theoretical
studies indicate that neighbouring galaxy clusters indeed appear to be aligned, with some
uncertainties due to the facts that shapes of clusters are notoriously hard to measure
observationally, and that the simulations used for the theoretical studies follow the
evolution of the dark matter, which may or may not trace the distribution of the galaxies
used to define observational cluster shapes.

The causes of such an alignment are not clear. Both infall of material (Van Haarlem \&
Van de Weygaert 1993) and tidal fields (Bond et al. 1996; also see Lee et
al. 2005a, 2005b) have been suggested as explanations. 

As shown in Colberg et al. (1999), the formation of clusters happens along filaments (also
compare the recent direct observations of this process for the $z=0.83$ cluster CLJ0152.7--1357 
in Maughan et al. 2005 and Tanaka et al. 2005). If the infall of material causes alignments
of clusters, then the presence of filaments is a prerequisite for alignments to exist.
We have investigated this scenario by studying the alignments of neighbouring clusters,
separating cluster pairs into those that are connected by a filament and those that are
not. For this study, we have made use of the filaments identified in Colberg et al. (2005).
As the rightmost column of Figure~\ref{fig:halo-fil-perpd-align} shows, there is a clear
difference between the two cases. Clusters connected by a filament are clearly aligned,
whereas unconnected ones are not. On the basis of this result it appears the Binggeli effect 
can be explained by the presence of filaments, along which material falls into the clusters.

Note that this result provides another indicator of the likely presence of filaments. As
Colberg et al. (2005) indicated, close pairs of clusters are candidates for the presence
of a filament in between them. The presence of alignment of the clusters adds another
strong indicator. However, given the difficulties involved when measuring cluster
shapes, this theoretical prediction might not be much help in practice.

It must be stressed that the presence of filaments does not mean that tidal fields
play no role whatsoever in the process of aligning the halos of galaxy clusters or of
galaxies themselves. As can be seen from the leftmost column of Figure~\ref{fig:halo-fil-perpd-align},
while filaments cause alignments of cluster--size halos, they have no discernible
influence on smaller halos. An alternative way to phrase this result is to say that 
halos of galaxies being aligned has nothing to do with whether those halos are embedded in
a filament or not. There also is no dependence of the alignments of galaxy size halos
on their distance from the nearest cluster, as shown in Figure~\ref{fig:halo-fil-projd-align}.

A somewhat esoteric question which one can ask is whether the strong alignments of halos
seen in simulations by e.g. Heavens et al (2000), Croft \& Metzler (2000) are due to 
halos being formed largely in filaments and being aligned with the filament direction.
In this case one would expect halos to be aligned with each other as a result. In this study,
we have found, however that halos are aligned with each other whether they fall in a filament
or not, so that this explanation is not valid. The tidal field around each galaxy sized halo
has a much more direct impact on the alignments of the halos.

We have investigated the contribution of tidal fields by modeling the tidal field of
a pair of clusters analytically as a simple model of two spherical masses. 
 Figure~\ref{fig:tidal-align} shows the 
alignments for the two tidal--field components $T_{zz}$ and 
$T_{\rho\rho}$. The stretching perpendicular to the cluster--cluster  axis is about twice 
that parallel to the cluster--cluster axis.  The magnitude of this effect is smaller than 
that of the alignment through filaments discussed above, but the abruptness of the
increase in alignment signal once tidal fields become substantial leads one to
the conclusion that tidal fields have a very direct influence on halo alignments.

In the light of the theoretical model of Lee et al. (2005a, 2005b), it is interesting to study
whether there exists a correlation of the magnitudes of
the ellipticities of halos with large-scale
structure. As we have already seen, the orientiations of the
ellipticities are clearly influenced by tidal
fields. Figure~\ref{fig:ellip-perpd} shows halo and cluster ellipticities as a function 
of the distance from cluster--cluster axes. There is no correlation between the 
ellipticities and the distance of the halos from the cluster--cluster axes,
whether filament or void. Furthermore, there is no difference 
between cluster pairs connected by a filament and those which have a void in between them.

We are thus left to conclude that while the alignment of clusters is dominated by the
infall of matter along filaments, for the vast majority of halos only tidal fields 
determine alignments of halos. As an aside, we note that 
this implies that Pimbblet (2005)'s
filament finding mechanism which is based on using alignments
to find filaments is unlikely to be successful. 
This judgment is however based on the assumption that the stellar parts of
galaxies will align in a similar way to the dark matter, something which must be tested.

\section*{Acknowledgments}

This project is supported by the National Science Foundation, NSF AST-0205978.

The simulation discussed here was carried out as part of the Virgo Consortium 
programme, on the Cray T3D/Es at the Rechenzentrum of the Max--Planck-Gesellschaft in 
Garching, Germany, and at the Edinburgh Parallel Computing Center. We are indebted to 
the Virgo Consortium for allowing us to use it for this work.

{}

\label{lastpage}

\end{document}